\documentclass{aa}  

\usepackage{graphicx}
\usepackage{xcolor}
\usepackage{bm}
\usepackage{enumitem}
\usepackage{subcaption} 
\usepackage{array}
\usepackage{amsmath}
\usepackage{amsfonts}
\usepackage{amssymb}
\usepackage{esint}
\usepackage{hyperref}
\hypersetup{
    colorlinks=true,
    linkcolor=black,
    citecolor=blue,
    urlcolor=magenta}
\usepackage{orcidlink}

\newcommand\calhoungrid[0]{disk grid}
\newcommand\Uqt[0]{\bm{U}}

\newcommand*{\bbar}[1]{\underline{\underline{#1}}}

\renewcommand{\div}{\bm{\nabla}\cdot}
\newcommand{\grad}{\bm{\nabla}}
\newcommand{\pdv}[2]{\frac{\partial #1}{\partial #2}}
\newcommand{\dv}[2]{\frac{\mathrm{d} #1}{\mathrm{d} #2}}
\newcommand{\dd}[1]{\,\mathrm{d}#1}
\newcommand{\abs}[1]{\left| #1 \right|}
\newcommand{\norm}[1]{\left\| #1 \right\|}

\title{Global solar compressible convection simulations\\ on non-Cartesian mappings}
\titlerunning{Global solar compressible convection simulations}

\author{Grégoire Doebele\inst{1}\corrauth{gregoire.doebele@sfr.fr}
    \and Allan Sacha Brun~\orcidlink{0000-0002-1729-8267}\inst{1}\email{sacha.brun@cea.fr}
    \and Maxime Delorme \inst{2}\email{maxime.delorme@cea.fr}
    }

\institute{Université Paris-Saclay, Université Paris Cité, CEA, CNRS, AIM, 91191, Gif-sur-Yvette, France
      \and Université Paris-Saclay, CEA, IRFU, 91191, Gif-Sur-Yvette, France
    }
    
\date{October 2026}

\begin{document} 

\abstract
   {Modeling the global dynamics of the Sun requires numerical frameworks capable of simulating the highly turbulent convective envelope and its coupling with the stable radiative interior. Simulating such spherical domains in standard Cartesian finite-volume codes presents significant topological challenges.}
   {The goal is to extend the Cartesian Adaptive Mesh Refinement (AMR) framework \texttt{Dyablo} to non-Cartesian geometries in order to perform global solar convection simulations that extend all the way down to the stellar core.}
   {We introduced a mapped grid approach within the \texttt{Dyablo} framework to build \texttt{Dyablo Whole Sun} (\texttt{DWS}) code. We implemented an $\alpha-\beta$ well-balanced scheme to preserve hydrostatic equilibrium. Furthermore, we present a robust methodology to generate fully integrated solar background profiles from $r=0$ to the surface.}
   {We validated the implementation using standard hydrodynamic tests and ran high-resolution 2D global solar convection simulations on a \textit{\calhoungrid{}}.}
   {Our models successfully capture the turbulent convective broad slow upflows and narrow fast downflows and the excitation of acoustic waves and p modes. The interaction of strong downflowing plumes with the top of the radiative zone, successfully excites internal gravity waves in the radiative interior.
   }


   \maketitle
   \nolinenumbers
   
\section{Introduction}

The Sun is a G2V-type star, and for this reason it possesses a highly turbulent and stratified outer convective envelope \citep{1996Sci...272.1286C,Brun2002Antia}, which leads to an intense time dependent complex surface dynamics. Generally, this surface convective turbulence, characterized by solar granulation with a typical size of 1000 km, is modeled in Cartesian coordinates and in local shallow boxes near the surface, in order to achieve the numerical resolution required to model it accurately \citep[][and references therein]{1998ApJ...499..914S,2005A&A...429..335V,2011A&A...531A.154G,2009LRSP....6....2N,2019A&A...631A.122K}.
However, the Sun’s dynamics also exhibit large-scale organization that cannot be easily captured by local approaches because of the need to consider curvature effects and deep domains. Consequently, over the past few decades, another type of HPC simulation of solar convection has been developed: global simulations in deep spherical shells \citep{1984JCoPh..55..461G,BRUN2004,2011Icar..216..120J,2021NatAs...5.1100H,2023SSRv..219...58K,Noraz2025}.
The drawback of this approach is that it is then difficult to resolve small scale dynamical features, such as solar surface granulation, as this requires very high resolution across the entire sphere, which is very costly in numerical resources if not adequately optimized. Further, given the very high stratification (with a large number of density scale heights in the solar convective envelope, e.g $N_{\rho,\odot} \simeq 13$) and the vastly different timescales between the surface ($\sim 5$ min) and the base of the convection zone at 200 Mm deep ($\sim 30$ days), one generally chooses not to include the outermost layers near the photosphere. Moreover, it is customary when modeling convection turbulence in spherical shells to apply Boussinesq-type fluid approximations \citep{1977GApFD...8...93G}, or, more appropriately for stars, the anelastic approach \citep{1969JAtS...26..448G,1981ApJS...45..335G,1986ApJS...61..585G,1995GApFD..79....1B,1999ApJS..121..247L,2011Icar..216..120J,2013ApJ...773..169V}, which filters out sound waves while imposing mass flux conservation, hence allowing larger time steps (CFL) with the added  constraint of considering low Mach numbers and high plasma $\beta$ conditions \cite[see also][for an alternative resolution of the fluid equations, using the reduced sound technique (RST)]{2015ApJ...798...51H}. 
However, all these fluid approximations have limitations when one wishes to model the near-surface convection, where flows become transonic, and its coupling to the stably stratified radiative interior via penetrative overshoot.
It was indeed shown by \cite{2012ApJ...756..109B}, that only the resolution of the fully compressible fluid equations perfectly captures the correct dynamics such as the excitation and propagation of gravity waves in the solar radiative interior.
It is therefore necessary to develop new methods to overcome these numerical and physical limitations, as well as to model the solar convective envelope in its entirety along with its coupling to a deep radiative interior and an atmosphere. The purpose of this article is to present an innovative approach based on the latest advances in numerical analysis and programming. In particular, we aim to use a fully compressible approach and variable resolution on a non-conforming grid, based on an adaptive mesh refinement method \citep{Berger1989, AMR2004} and an original mathematical solution for handling spherical geometry to lay down a high resolution model of global solar turbulent convection with an optimal use of numerical resources.

In this study, we lay down the groundwork for the \texttt{Dyablo} framework \citep{Delorme2025} (appendix~\ref{anx:dyablo}) and its Whole Sun branch (or \texttt{DWS}). We will first explain how we will modify the main loop in \texttt{DWS} to apply a non Cartesian mapping  transformation from the Cartesian grid to deformed grids based on \cite{COLELLA20112952}, and then to extend it to solve the fluid equations on disk or sphere-like grid configurations, more suited to model stars, based on \cite{calhoun2008}. We will validate the implementation by performing standard unit tests (such as Riemann-2D test), followed by Rayleigh-Taylor instability tests to account for the gravity required in convection problems. Next, we develop convection setups based on proven configurations such as \cite{Hurlburt1986} but using a ring grid. 
Finally, we develop a fully compressible spherical model of solar convection with mesh refinement and perform various tests: First, numerical tests on the convergence and accuracy of the solutions, and Second, on the influence of control parameters such as $N_\rho$ or the Prandtl or Rayleigh  numbers $\mathrm{Pr}$, and $\mathrm{Ra}$. Note that in this first study of solar convection, we do not include rotation, which will be addressed in a future study. Instead, we focus on non rotating global convection properties and its associated heat transport, in the spirit of the study done by \cite{2016ApJ...818...32F}, by studying the influence of increasing the degree of turbulence and the realism of the modeled turbulent convection plumes.

The paper is organized as follows. In Section 2, we detail the numerical methods for our finite volume approach, including the grid mapping, the flux computation, and the implementation of a well-balanced scheme for irregular grids. In Section 3, we validate our framework against standard 2D hydrodynamic benchmarks and present the background state and the results of the dynamical properties of our global 2D solar convection model. In section 4, we reflect upon our results and conclude.

\section{Numerical methods}
\subsection{Governing equations}
To model solar convection, we solve the compressible Navier-Stokes equations. We consider the compressible formulation because we wish to model the solar surface -- where the compressibility cannot be ignored -- as well as to track the acoustic waves and connect to helioseismology. In conservative form, the compressible fluid equations read:

\begin{equation}
\begin{split}
    \label{eq:navier_stokes1}   
    &\pdv{\rho}{t} + \div{(\rho \bm{u})} = 0 \\
    &\pdv{\rho \bm{u}}{t} + \div{(\rho \bm{u} \otimes \bm{u} + p\bbar{\mathbb{I}})} = \div{\bbar{\sigma}} + \rho \bm{g} \\
    &\pdv{E}{t} + \div{((E+p)\bm{u})} = \div{(\bbar{\sigma}\cdot \bm{u})} + \div{(K\grad{T})} \\
    &\phantom{\pdv{E}{t} + \div{((E+p)\bm{u})} = } + \rho \, \bm{g}\cdot\bm{u} + q
\end{split}
\end{equation}

The system is closed by the state equation of the internal energy $\varepsilon = \frac{1}{\gamma - 1} \frac{p}{\rho}$, and by the ideal gas equation $p = R\rho T$, with $R$ the specific gas constant. The total energy density is defined by $E = \rho \varepsilon + \frac{1}{2}\rho \norm{\bm{u}}^2$, with 
density ($\rho$),
velocity vector ($\bm{u}$),
pressure ($p$),
total energy density ($E$),
temperature ($T$),
thermal conductivity ($K$),
viscous stress tensor ($\bbar{\sigma}$),
gravitational acceleration ($\bm{g}$),
internal heating ($q$),
identity matrix ($\bbar{\mathbb{I}}$).

\subsection{Grid}

For global solar physics simulations, we require a computational domain with a spherical geometry. While polar or spherical coordinates are natural choices for such geometries, they are not natively possible in \texttt{Dyablo} as it is a finite-volume code built for Cartesian Adaptive Mesh Refinement (AMR) grids. 
Furthermore, changing the coordinate system would introduce geometrical source terms, coordinate singularities (at the center and along the $z$-axis), and dealing with degenerate cells in the neighboring of those singularities. These highly distorted cells severely restrict the simulation time step, which is proportional to the smallest cell size on the whole domain. 

\begin{figure}
    \centering
    \includegraphics[width=0.75\linewidth]{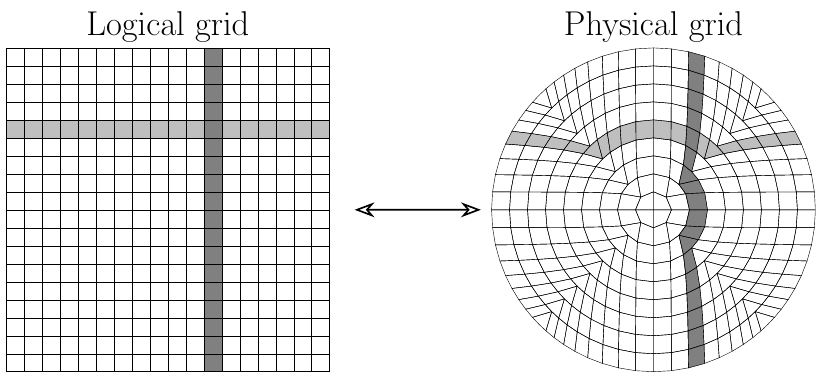}
    \caption{Mapping from the logical Cartesian grid (left) to the physical \calhoungrid{} (right). The colored row and column correspond to the same cells on both the Cartesian and physical grids.}
    \label{fig:map_calhoun}
\end{figure}

\begin{figure}
    \centering
    \includegraphics[width=0.7\linewidth]{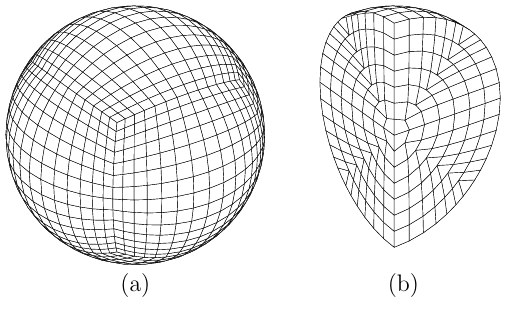}
    \caption{Mapping for the 3D spherical grid as a generalization of the 2D \calhoungrid{} (Fig.~\ref{fig:map_calhoun}). (a) Surface of the grid, (b) Cross-section view of a quarter of the grid.}
    \label{fig:map_calhoun_3d}
\end{figure}

To circumvent these limitations, we use a mapping approach using a bijection that transforms each vertex from a Cartesian \textit{logical grid} ($\xi, \eta, \zeta$) to a \textit{physical grid} ($x, y, z$). Following the work of \cite{calhoun2008, Calhoun2010}, we implement a mapping that deforms a logical Cartesian square (or cube) into a physical disk (or sphere) (Fig.~\ref{fig:map_calhoun}; Fig.~\ref{fig:map_calhoun_3d}). 

This specific mapping is characterized by its radial consistency: it maps points with the same Chebyshev distance $\max{(\abs{\xi}, \abs{\eta})}$ in the logical grid on points with the same radii in the physical grid. Mathematically, for a 2D domain $(\xi, \eta) \in [-1, 1]^2$, the \calhoungrid{} mapping is defined as:
\begin{equation}
    \begin{pmatrix} x \\ y \end{pmatrix} = R\, \frac{\max(\abs{\xi},\abs{\eta})}{\sqrt{\xi^2 + \eta^2}} \, \begin{pmatrix} \xi \\ \eta  \end{pmatrix} 
\end{equation}
where $R$ is the surface radius. The radius is defined by $r = \sqrt{x^2 + y^2}$. 

An important advantage of this grid is its relatively low size difference across cells of the domain: the ratio between the largest and smallest cell volumes remains approximately $2$, preventing extreme CFL restrictions. However, a drawback is the distortion along the diagonals caused by the non-differentiability of the mapping. In these regions, cells become highly non-orthogonal (approaching triangular shapes) and do not share faces with cells at the immediate lower radial positions. 
This distortion can lead to numerical artifacts and a localized reduction in the accuracy of the solution, an effect particularly pronounced in highly stratified environments such as solar simulations. To mitigate these drawbacks, we use a well-balanced scheme to maintain hydrostatic stability, combined with an unstructured-type reconstruction based on a gradient-based method \citep{berberich2019}.

Other alternative methods for spherical grid include:
\begin{itemize}
    \item Star-in-a-box: The sphere is embedded in a large Cartesian cube. This is easy to implement but requires complex inner-boundary conditions at the stellar surface and present large unused regions outside the star \citep{Dobler2006, Kapyla2021, Steffen2007}.
    \item Cubed-Sphere: This method glues six (or seven if we include the center) independent curvilinear patches (the `net' of a cube) to form a shell. It offers better cell regularity and avoids the diagonal distortions mentioned above. However, it is significantly more complex to integrate into an AMR framework, as it requires managing communications and interpolation across six different grid orientations \citep{Ivan2011, Colella2016}. 
    \item Yin-Yang grid: This approach decomposes the spherical domain into two identical, overlapping latitude-longitude grid patches \citep{Kageyama2004}. While it successfully avoids polar singularities and provides good local cell regularity, it relies on an overlapping grid strategy. Consequently, the cells do not match perfectly at the boundaries, which requires high-order interpolation in the overlapping regions, and makes implementing AMR particularly challenging.
\end{itemize}

In this work, we adopt the single deformed grid method for its balance of simplicity and compatibility with \texttt{Dyablo}'s AMR. 
For validation purposes, we introduce two curvilinear grids: a deformed Cartesian grid \citep{COLELLA20112952, PERSSON20091585} and a ring grid.
The deformed Cartesian grid (an example of can be seen in figure~\ref{fig:lax_deform}) is used to verify the solver through 2D unit tests and via direct comparison with the original solver on a standard Cartesian grid. This grid is defined by:
\begin{equation}
     \begin{pmatrix} x\\y \end{pmatrix} = \begin{pmatrix} \xi\\\eta \end{pmatrix} + \begin{pmatrix} c_\xi\\c_\eta \end{pmatrix} \,\sin(2\pi \xi)\sin(2\pi \eta)
\end{equation}
the grid is 1-periodic in $\xi$ and $\eta$, with $c_\xi$ and $c_\eta$ deformation factors along both directions. These factors have to be in $[0, \frac{1}{2\pi}[$ for the mapping to be injective.

Similarly, the ring grid maps $(\xi, \eta) \in [0, 1]^2$ to the physical coordinate with: 
\begin{equation}
    x = \eta \cos(-2\pi\xi), \quad y = \eta \sin(-2\pi\xi)
\end{equation}
where $\eta \in [r_\mathrm{in}, r_\mathrm{out}]$. An example can be seen in the figure~\ref{fig:rt_ring_evol}.

\subsection{Geometrical properties}

\begin{figure}
    \centering
    \includegraphics[width=0.45\linewidth]{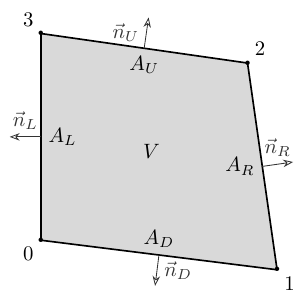} 
    \caption{Geometric notations for a 2D cell. The cell volume $V$ denotes the 2D area, while $A_f$ and $\bm{n}_f$ represent the face areas (edge lengths) and outward unit normals, respectively, for each direction (Left, Right, Down, Up). Vertices are indexed in counter-clockwise order.}
    \label{fig:cell_2d_schema}
\end{figure}

While the framework is designed to support full sphere 3D setups, the present paper focuses more specifically on 2D global simulations. In our implementation, physical cells are treated as general quadrilaterals. In the following, we will use the 3D terminology for geometrical properties. For a 2D cell defined by vertices $\{\bm{r}_i\}_{0 \le i \le 3}$ (Fig.~\ref{fig:cell_2d_schema}), the volume $V$ (corresponding to the 2D cell area), the centroid and the face area $A_f$ (corresponding to the 2D edge lengths) are computed as follows:\paragraph{2D case:}
\begin{itemize}
    \item Volume: 
        \begin{equation*}
            V = \frac{1}{2} \abs{(x_0 - x_2) (y_1 - y_3) - (y_0 - y_2) (x_1 - x_3)}
        \end{equation*}
    \item Face area: 
        \begin{equation*}
            A = \norm{\bm{r}_i - \bm{r}_{i+1}}
        \end{equation*}
    \item Centroid: 
    \begin{equation*}
        \begin{split}
          \begin{pmatrix}x\\y\end{pmatrix} = \frac{1}{3V} \sum_{i=0}^3 \begin{pmatrix}
          x_i + x_{i+1} \\
          y_i + y_{i+1}
          \end{pmatrix} (x_i y_{i+1} - x_{i+1} y_i)
        \end{split}
    \end{equation*}
    where vertices are ordered counter-clockwise, and $\bm{r}_4$ is identified to $\bm{r}_0$.
    \item Rotation matrix:
        \begin{equation*}
            R = \begin{pmatrix}
            n_x & -n_y \\
            n_y & n_x
            \end{pmatrix}
        \end{equation*}
\end{itemize}

The computation of the geometrical properties for the 3D case is detail in appendix~\ref{anx:geo3d}. 

\subsection{Discretization}
Now that the mapping and the geometrical properties of cells is established, we can describe the numerical methods used to solve the governing conservation laws.
Hyperbolic terms (hydrodynamics) and parabolic terms (thermal conduction and viscosity) are treated separately. The governing equations are spatially discretized using the finite volume method \citep{leveque_2002}.

    \subsubsection{Hydrodynamic system of equation}
    We first consider the 2D system of compressible Euler equations, which models the conservation of mass, momentum, and energy for a fluid system. Let $\Uqt$ be the state vector of conservative quantities and $\bm{F}(\Uqt)$ the flux function:
    \[
        \Uqt = \begin{pmatrix} \rho \\ \rho u \\ \rho v \\ E \end{pmatrix}^\top,
        \qquad
        \bm{F}(\Uqt) = 
        \begin{pmatrix} F(\Uqt) \\ G(\Uqt) \end{pmatrix}
    \]
    where $F(\Uqt)$ and $G(\Uqt)$ are the flux in $x$ and $y$ direction:
    \[
        F(\Uqt) = \begin{pmatrix}  \rho u \\ \rho u^2 + p \\ \rho vu           \\ (E+p)u\end{pmatrix}^\top, \qquad
        G(\Uqt) = \begin{pmatrix}  \rho v \\ \rho uv      \\ \rho v^2 + p      \\ (E+p)v\end{pmatrix}^\top
    \]

    In its conservative form, the Euler system with gravitational source terms is written as:
    \begin{equation}
        \label{eq:euler}
        \pdv{\Uqt}{t} + \div{\bm{F}(\Uqt)} = S(\Uqt), \qquad S(\Uqt) = \begin{pmatrix}  0 \\ \rho g_x \\ \rho g_y  \\ \rho \bm{g} \cdot \bm{u}\end{pmatrix}^\top
    \end{equation}
    where $S(\Uqt)$ is the gravitational source terms.

    Using the finite volume method \citep{leveque_2002} and Godunov's scheme \citep{Godunov1959} to compute fluxes via a Riemann solver \citep{Toro2009}, we obtain the following spatial discretization for a given cell $i$:
    \begin{equation}
        \label{eq:volumesfinis3}
        \Uqt^{n+1}_i = \Uqt^n_i - \frac{\Delta t}{V_i} \sum_{f \, \in \, \text{faces}} A_f \, \Hat{\bm{F}}_f
    \end{equation}
    where $\Uqt^n_i$ is the vector of conservative quantities averaged over the cell $i$ at time step $n$, and $V_i$ is the cell volume. $A_f$ is the area of the face $f$, and $\Hat{\bm{F}}_f$ is the numerical flux per unit area passing through that face. The method for calculating these numerical fluxes will be detailed in section \ref{sec:hydroflux}. The time integration scheme and time step constraints are detailed in appendix~\ref{anx:timescheme}.

    \subsubsection{Thermal conduction}
    To extend the ideal hydrodynamic description of the Euler equations to realistic stellar environments, non-adiabatic and dissipative processes must be included. To complete the description, radiative transfer inside the solar interior is treated under the radiative diffusion approximation, which reduces the radiative transport to a simple diffusion process governed by the heat equation:
    \begin{equation}
    \begin{split}
        \label{eq:eq_chaleur}
        \pdv{E}{t} - \div{(K\grad{T})} = 0
    \end{split}
    \end{equation}
    
    After integrating over the cell volume and applying the divergence theorem, the semi-discretized form is:
    \begin{equation}
        \label{eq:volumesfinis_TC}
        \pdv{E_i}{t} = \frac{1}{V_i} \sum_{f \, \in \, \text{faces}} (K \grad{T})_f \, \cdot \bm{A}_f 
    \end{equation}
    where $E_i$ is the total energy averaged over the cell $i$. The temperature gradient $(K \grad{T})_f$ is evaluated at the center of the cell faces, and $\bm{A}_f$ is the outward-pointing normal area vector of the face.
    
    \subsubsection{Viscosity}
    Along with thermal diffusion, the fully compressible Navier-Stokes system requires including viscous transport. The parabolic terms related to momentum and energy dissipation due to fluid friction are given by:
    \begin{equation}
    \begin{split}
        \label{eq:eq_visco}
        \pdv{\rho \bm{u}}{t} &- \div{\bbar{\sigma}} = 0 \\
        \pdv{E}{t} &- \div{(\bbar{\sigma} \cdot \bm{u})} = 0 \\
    \end{split}
    \end{equation}
    where $\bbar{\sigma} = \mu\, (\grad{\bm{u}}  + \grad{\bm{u}}^\top - \frac{2}{3} (\div{\bm{u}}) \bbar{I}\,)$ is the viscous stress tensor for a Newtonian fluid, and $\mu$ is the dynamic viscosity. 
    
    Similarly, the spatial discretization yields:
    \begin{equation}
    \begin{split}
        \label{eq:volumefinis_visco}
        \pdv{(\rho \bm{u})_i}{t} &= \frac{1}{V_i} \sum_{f \, \in \, \text{faces}} \bbar{\sigma}_f \cdot \bm{A}_f  \\
        \pdv{E_i}{t} &= \frac{1}{V_i} \sum_{f \, \in \, \text{faces}} (\bbar{\sigma} \cdot \bm{u})_f \cdot \bm{A}_f
    \end{split}
    \end{equation}

\subsection{Computation of fluxes}
The calculation of numerical fluxes at the interfaces is a two-step process: first, the reconstruction of states on both sides of the interface, followed by the resolution of a local Riemann problem.

    \subsubsection{Reconstruction of states}
    To achieve second-order spatial accuracy, the fluid states are reconstructed at each cell interface. While standard Cartesian configurations typically rely on one-dimensional slope limiters, e.g. Piecewise linear method (PLM) \citep{vanLeer1979, leveque_2002}, non-Cartesian or mapped grids require a more general, multi-dimensional approach to account for the grid's non-orthogonality.
    
    In \texttt{Dyablo}, this reconstruction is performed on the primitive variables $\bm{Q} = (\rho, u, v, p)^\top$, as they are significantly less prone to spurious oscillations than their conservative counterparts. The cell-centered gradient $\grad{\bm{Q}}$ is computed using the least squares method with a stencil containing all adjacent and diagonal cells. The reconstructed state $\bm{Q}_f$ at the center of a face $f$ is then given by:
    \begin{equation}
        \bm{Q}_f = \bm{Q}_c + \grad{\bm{Q}} \cdot \bm{r}_{c \to f}
    \end{equation}
    where $\bm{Q}_c$ is the primitive state at the cell center and $\bm{r}_{c \to f}$ is the vector pointing from the cell center to the face center.

    In Dyablo we have access to various types of limiters, which have been adapted for non-Cartesian mapping following \citep{Darwish2003}, and are being used to limit all the primitive variables. In this study, we use the TVD-MUSCL scheme with a minmod limiter and the `exact r formulation'.

    \subsubsection{Hyperbolic flux}
    \label{sec:hydroflux}
    The evaluation of hyperbolic fluxes across a surface is treated as a local Riemann problem at the interface. We use the HLLC Riemann solver (Harten-Lax-van Leer-Contact) \citep{Toro1994}, which offers a good balance between numerical robustness and the accurate resolution of contact discontinuities.
    
\begin{figure}
        \centering
        \includegraphics[width=0.9\linewidth]{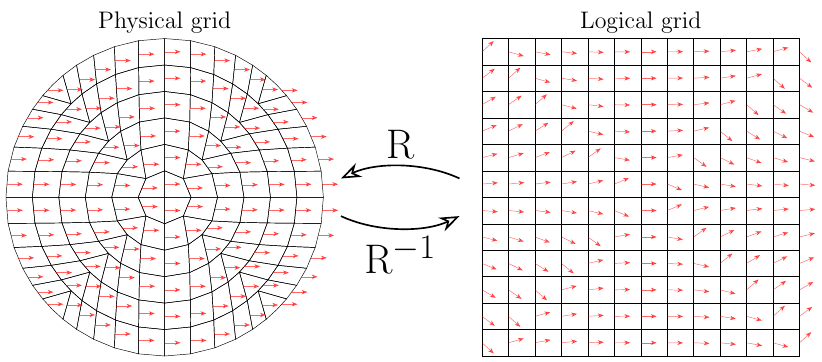}
        \caption{Transformation of a unidirectional flow (to the right) from the physical grid to the logical grid. Only vertical interfaces are considered.}
        \label{fig:rotate_map_flux}
    \end{figure}

    To apply this solver to our mapped grid, we follow the methodology proposed by \cite{Toro2009}. The process involves a local coordinates transformation:
    \begin{enumerate}
        \item Rotation to local frame: Before solving the Riemann problem, velocity vectors and other directional quantities (e.g. magnetic field) are rotated from the physical space to a local logical frame aligned with the face normal (Fig.~\ref{fig:rotate_map_flux}). Scalar quantities remain unchanged.
        \item Riemann solver: The flux across the interface is computed in this 1D local Cartesian frame using the standard HLLC formulas.
        \item Inverse rotation: The resulting numerical flux is then rotated back into the physical coordinates system to be used in the final cell update.
    \end{enumerate}
    This approach allows us to use the Cartesian Riemann solvers while maintaining the geometric integrity of the non-Cartesian simulation. 
    
    \subsubsection{Diffusive terms}
    The computation of parabolic terms (thermal conduction and viscosity) requires an accurate estimation of the gradients at the face center. 
    Because the center-to-center vector does not intersect the shared face at its centroid on mapped grids, a standard finite difference approximation introduces significant geometric errors.
    
    To resolve this issue, we implement the \textit{over-relaxed correction} approach \citep{Jasak1996}. This technique decomposes the gradient into a primary component along the line connecting cell centers and a secondary cross-diffusion correction term. This ensures that the diffusive fluxes remain consistent and stable even in regions of the grid where cells are significantly distorted (such as the one along the diagonals of the \calhoungrid{}). It is given by:
    \begin{equation}
        (\grad{\bm{Q}})_f = \overline{(\grad{\bm{Q}})}_f + \frac{\bm{Q}_R - \bm{Q}_L}{\norm{\bm{r}_{LR}}}\bm{e}_{LR} - \left(\overline{(\grad{\bm{Q}})}_f \cdot \bm{e}_{LR}\right) \bm{e}_{LR}
    \end{equation}
    with $\bm{r}_{LR}$ the vector between the centers of the left cell and the right cell. $ \bm{e}_{LR}$ is the unit vector of $\bm{r}_{LR}$. The term $\overline{(\grad{\bm{Q}})}_f$ is an approximation of the face-centered gradient obtained by an average of the two face-centered gradients on the face, such as in \citep{Moukalled2016}:
    \begin{equation}
        \overline{(\grad{\bm{Q}})}_f = g_L\, (\grad{\bm{Q}})_L + (1-g_L)\, (\grad{\bm{Q}})_R 
    \end{equation}
    where the factor $g_L = \frac{\bm{r}_{Lf}\cdot \bm{e}_{LR}}{\norm{\bm{r}_{LR}}}$.
    
    \subsubsection{Well-balancing}
    \label{sec:wellbalacing}
    A numerical scheme is said to be \textit{well-balanced} if it is capable of preserving a specific steady state, most commonly the hydrostatic equilibrium, exactly at the discrete level in the presence of source terms \citep{Greenberg1996}. 
    In the context of solar convection, where the pressure gradient and gravitational force are nearly in balance, a well-balanced scheme is essential to suppress spurious numerical velocities inherent to the finite volume discretization, preventing them from polluting the simulation, especially on non-regular or mapped grids \citep{CaniveteCuissa2022}.
    
    Implementing a general well-balanced scheme for arbitrary configurations is technically challenging. For this work, we adopted the $\alpha$--$\beta$ method proposed by \cite{berberich2019}. While less general than some alternative methods \citep{Kappeli2016}, this approach is highly effective and particularly well-suited for any grids. It relies on reconstructing a normalized state using the analytical profiles of the initial hydrostatic solution. 
    
    Let $\alpha(r)$ and $\beta(r)$ denote respectively the analytical radial profiles for density and pressure, whose derivations are detailed in section \ref{sec:backgroundstate}. We utilize these profiles to reconstruct the left and right states at an interface. We first apply a transformation $\mathcal{T}$ to the primitive variables $\bm{Q}$ to obtain the reduced state $\tilde{\bm{Q}}$:
    \begin{equation}
        \bm{Q} = \begin{pmatrix}\rho\\u\\v\\p\end{pmatrix}^\top \xrightarrow[\quad\mathcal{T}\quad]{} \tilde{\bm{Q}} = \begin{pmatrix}{\rho}/{\alpha}\\u\\v\\{p}/{\beta}\end{pmatrix}^\top
    \end{equation}
    A spatial reconstruction is then performed on the reduced variables $\tilde{\mathbf{Q}}$. Finally, the inverse transformation $\mathcal{T}^{-1}$ is applied to the reconstructed interface values to recover the well-balanced primitive states. For example, the reconstruction of the left-state density $\rho_L$ at an interface $i+\frac{1}{2}$ is computed as:
    \begin{equation}
        \rho_i \xrightarrow{\mathcal{T}} \tilde{\rho}_i = \frac{\rho_i}{\alpha(r_i)} \xrightarrow[(\grad{\rho})_i]{\text{recons.}} \tilde{\rho}_{L} = \tilde{\rho}_{i+\frac{1}{2}} \xrightarrow{\mathcal{T}^{-1}} \rho_L = \alpha(r_{i+\frac{1}{2}}) \tilde{\rho}_L
    \end{equation}
    
    To ensure consistency, the discrete gravitational source term $\mathbf{g}_{i}$ must be formulated to exactly cancel the discrete pressure gradient of the background state. Following \cite{berberich2019}, we use:
    \begin{equation}
        \bm{g}_{i} = \frac{1}{\alpha(r_{i})\,V_i} \sum_{f \in \text{faces}} \beta(r_f) \bm{A}_f
    \end{equation}
    where $\bm{A}_f$ is the outward-pointing area vector of face $f$. The momentum and energy updates are then given by:
    \begin{equation}
        \pdv{(\rho \bm{u})_i}{t} = \rho_i \bm{g}_i, \qquad
        \pdv{E_i}{t} = \rho_i \bm{g}_i \cdot \bm{u}_i
    \end{equation}
    We show the validity of this well-balanced method in Fig.~\ref{fig:wb_test_hydrostatic}.

    Finally, this well-balancing strategy is extended to the computation of temperature gradients in the thermal conduction module. By defining the background temperature profile $\Gamma = \beta / (\alpha R)$, we can decompose the temperature gradient using Leibniz's rule:
    \begin{equation}
        (\grad{T})_f = \Gamma_f \left(\grad{\frac{T}{\Gamma}} \right)_f + \left(\frac{T}{\Gamma} \right)_f (\grad{\Gamma})_f
    \end{equation}
    Since $\Gamma$ and $\grad{\Gamma}$ are known analytically, and the terms $(T/\Gamma)_f$ and $\grad{(T/\Gamma)}_f$ are obtained through the reconstruction of reduced variables, this formulation improves the precision of the diffusive fluxes.

    \subsubsection{Boundary conditions}
    The \calhoungrid{} was introduced to avoid a singularity at the center, hence only outer boundary conditions are required at the surface of the computational domain.
    At this boundary, ghost cells are populated to maintain the background hydrostatic equilibrium for density and pressure. For the velocity field, we apply stress-free impenetrable boundary conditions:
    \begin{equation}
         v_r^\mathrm{face} = 0, \qquad \dv{}{r} \left( \frac{\bm{v}_\perp^\mathrm{face}}{r} \right) = \bm{0}
    \end{equation}

    To populate the ghost cells (subscript $\mathit{g}$) from the inner adjacent cells (subscript $\mathit{in}$), we rely on the $\alpha$--$\beta$ analytical profiles for the thermodynamic variables (density, pressure). We use a finite difference approach to discretize the velocity constraint of the previous equation. The ghost cell states are thus reconstructed as:
    \begin{equation}
    \begin{split}
        \rho_\mathrm{g} &= \rho_\mathrm{in}\,\frac{\alpha_\mathrm{g}}{\alpha_\mathrm{in}}, \qquad p_\mathrm{g} = p_\mathrm{in}\,\frac{\beta_\mathrm{g}}{\beta_\mathrm{in}} \\
        \bm{v}_\mathrm{g} &= \bm{v}_\mathrm{in} \frac{r_\mathrm{g}}{r_\mathrm{in}} - \left(\frac{r_\mathrm{g}}{r_\mathrm{in}} - \frac{r_\mathrm{face} - r_\mathrm{g}}{r_\mathrm{face} - r_\mathrm{in}}\right) (\bm{v}_\mathrm{in} \cdot \bm{e}_r) \bm{e}_r
    \end{split}
    \end{equation}

    Finally, the thermal boundary condition is set by fixing the radial temperature gradient at the surface to match the background stratification:
    \begin{equation}
        \left. \dv{T}{r} \right|_\mathrm{face} = \dv{\Gamma}{r}{}(r_\mathrm{face})
    \end{equation}
    where $\Gamma = \beta / (\alpha R)$ is the analytical background temperature defined by the $\alpha$--$\beta$ well-balancing scheme.

\subsection{Well-balanced adaptive mesh refinement}
    \texttt{Dyablo} features block-based Adaptive Mesh Refinement (AMR) capabilities \citep{Berger1989, AMR2004}. The computational domain is decomposed into multiple independent blocks, where a maximum difference of one refinement level is allowed between adjacent blocks. 
    Each block is surrounded by a layer of ghost cells with the same refinement level as the patch itself. At a refinement transition boundary, these ghost cells must be populated using data from neighboring patches at a different resolution.
    Standard AMR interpolation operators assume a uniform state within cells, which breaks the hydrostatic equilibrium in stratified stellar interiors. To preserve this equilibrium across refinement jumps, we directly incorporate the analytical background profiles $\alpha(r)$ and $\beta(r)$ into the mesh transition operators:
    
    \begin{itemize}
        \item Fine-to-Coarse: Fine cells velocities are volume-averaged onto the coarse ghost cell. For the stratified states, a well-balanced volume-weighted average is performed to define the coarse density $\rho_\text{big}$ and pressure $p_\text{big}$:
        
        \begin{equation*}
            \rho_\text{big} = \frac{1}{V_\text{big}} \sum_i V_i \rho_i \frac{\alpha(r_\text{big})}{\alpha(r_i)}
        \end{equation*}
        The coarse pressure $p_\text{big}$ is obtained analogously by replacing $\alpha(r)$ with the pressure profile $\beta(r)$.
        
        \item Coarse-to-Fine: Velocities are directly mapped onto fine cells. Density and pressure states are set by scaling the coarse cell state using the analytical profiles:
        \begin{equation*}
            \rho_i = \rho_\text{big} \frac{\alpha(r_i)}{\alpha(r_\text{big})}
        \end{equation*}
        The fine pressure $p_i$ is similarly scaled using $\beta(r)/\beta(r_\text{big})$.
    \end{itemize}

\begin{figure}
        \centering
        \includegraphics[width=0.8\linewidth]{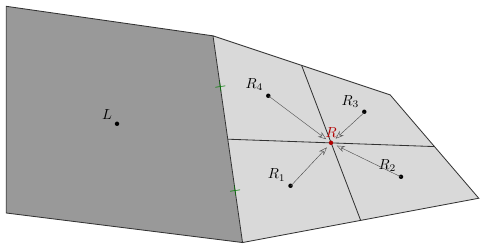}
        \caption{Reconstruction of a coarse cell center from fine cells centers at level transitions.}
        \label{fig:cell_amr}
    \end{figure}

    To maintain strict global conservation at refinement transitions, the numerical fluxes must be perfectly consistent across the shared faces. In practice, the fine cells compute the accurate fluxes across the boundary. These fine fluxes are then summed over the shared face area to update the adjacent coarse cell, replacing the coarse cell's own lower-resolution flux calculation at that specific interface. (e.g., in Figure \ref{fig:cell_amr}, the fine cells $R_1$ and $R_4$ determine the total flux that will update the coarse cell through its right interface).
    
    Finally, we must ensure consistency between the flux calculation and the source term integration at the shared interface. Just as the coarse cell receives its flux update from the adjacent fine cells, it must also receive the matching discrete gravity contribution from that specific face $f$ to remain perfectly balanced. The gravity contribution on the coarse cell from the shared interface is explicitly evaluated using the coarse cell's own state ($\rho_{\text{big}}$, $\bm{u}_\text{big}$) combined with the face-centered analytical pressure profile:
    \begin{equation*}
        \bm{g}_{\text{big},f} = -\frac{\rho_\text{big}}{\alpha(r_\text{big})} \frac{1}{V_\text{big}} \beta(r_f) \bm{A}_f
    \end{equation*}
    This rigorous flux and source-term matching across resolution jumps guarantees that the $\alpha$--$\beta$ well-balanced scheme is globally preserved throughout the AMR hierarchy.

\section{Results}

In this section, we validate the \texttt{DWS} solvers on mapped grids through benchmarks of increasing physical complexity: a 2D Riemann test (pure hydrodynamics on the deformed Cartesian grid), Rayleigh-Taylor instabilities (introducing gravity on the ring grid and \calhoungrid{}), a polytropic convection test on the ring grid. Then we will show and analyze a global solar convection simulation.

\subsection{2D Riemann test}
    The 2D Riemann tests, extensively categorized by \cite{laxliu:tests}, serve as benchmarks for validating multidimensional hydrodynamic solvers. These setups, which generalize the classic 1D Sod shock tube, divide a square domain into four quadrants, each initialized with a constant primitive state $\bm{Q} = (\rho, u, v, p)^\top$. Because these tests are highly sensitive to numerical dissipation and grid alignment, they are ideal for evaluating solver robustness on deformed geometries. In this work, we focus on Configuration 3 from \cite{laxliu:tests}, which is one of the most sensitive and has extremely sharp wavefronts.
    
\begin{figure*}
    \centering
        \begin{tabular}{ccccc}
            \includegraphics[width=0.18\linewidth]{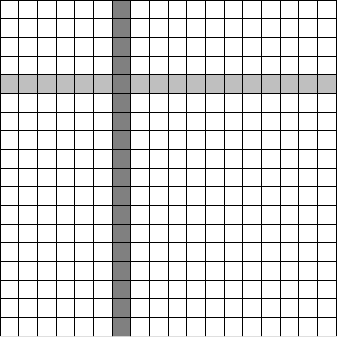} &
            \includegraphics[width=0.18\linewidth]{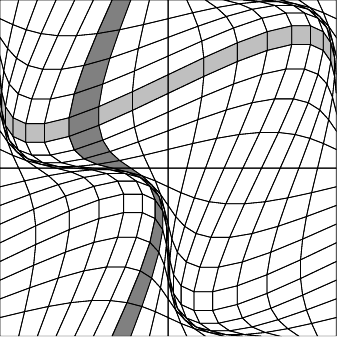} &
            \includegraphics[width=0.18\linewidth]{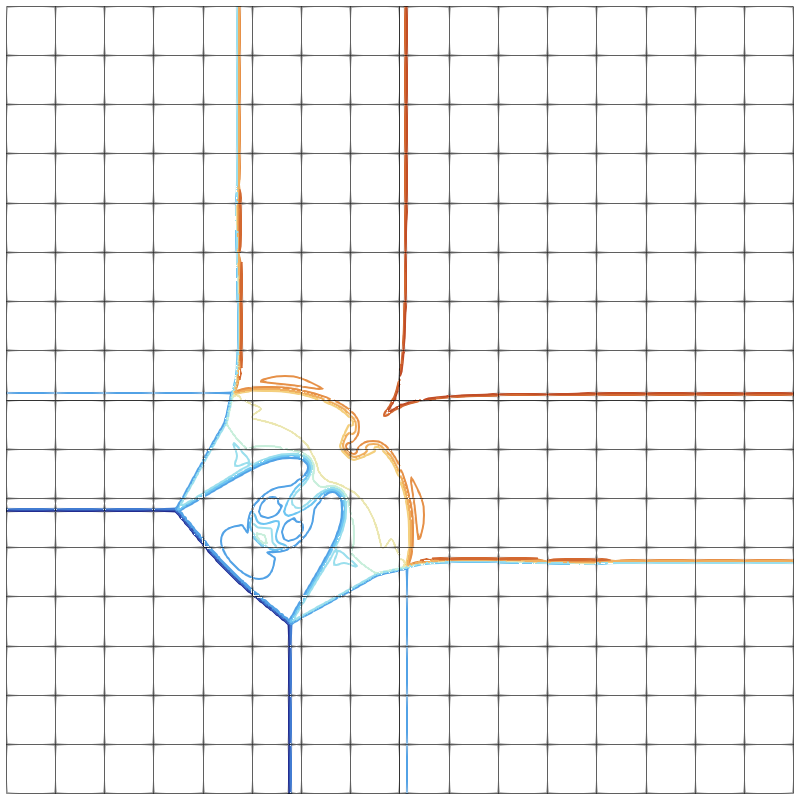} &
            \includegraphics[width=0.18\linewidth]{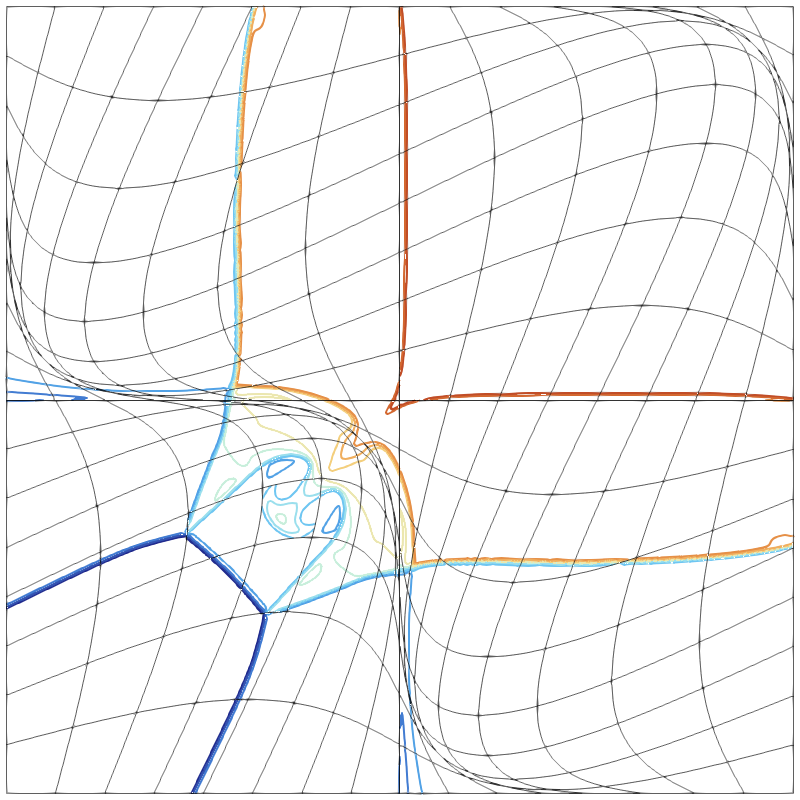} &
            \includegraphics[width=0.18\linewidth]{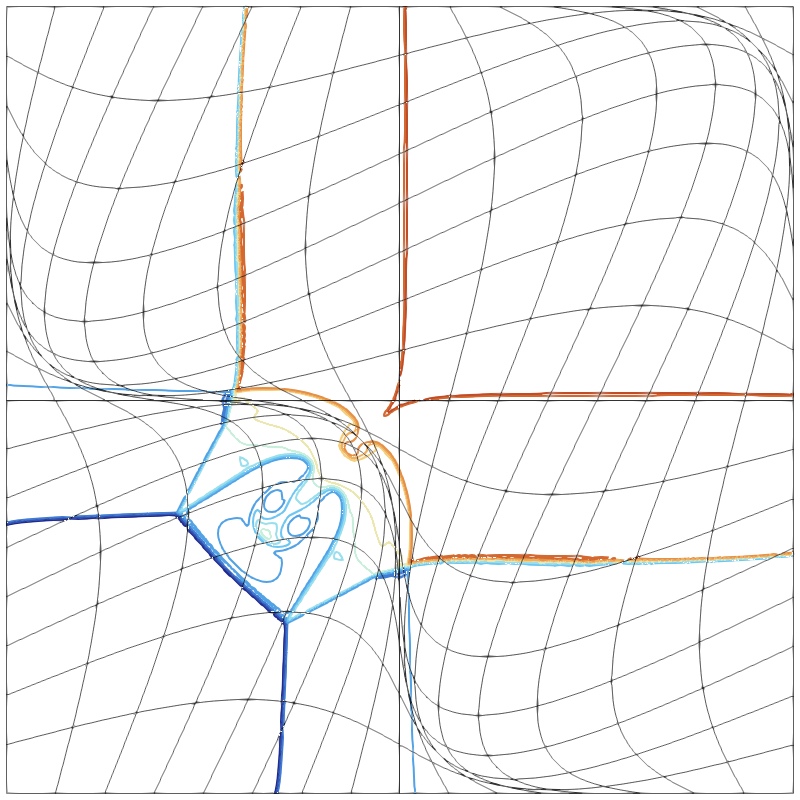} \\
            \small (a) & \small (b) & \small (c) & \small (d) & \small (e)
        \end{tabular}
        \caption{Demonstration of the boundary condition fix on a $512^2$ deformed Cartesian grid using the \cite{laxliu:tests} configuration 3 benchmark (2D Riemann test). (a) Cartesian grid. (b) Highly deformed Cartesian grid with a deformation coefficient of $c=0.15$ (the maximum is $1/2\pi \approx 0.159$). (c) Reference density contours on the Cartesian grid. (d) Deformed Cartesian grid without boundary fix: the standard ghost-cell extrapolation enforces a zero gradient along the deformed logical lines rather than the physical normals (Fig.~\ref{fig:laxliu_interpolation}), causing the wavefront bending. (e) Deformed grid run with the boundary fix, which recovers the correct physical behavior.}
        \label{fig:lax_deform}
    \end{figure*}

    On a standard uniform Cartesian grid, \texttt{DWS} accurately reproduces the reference solution (Fig.~\ref{fig:lax_deform}c). However, when executing the same test on the deformed grid, we observe a curvature of the wavefronts near the domain boundaries. This artifact originates from the standard implementation of zero-gradient outflow boundary conditions on non-orthogonal meshes.
    
    In a pure Cartesian framework, a zero-gradient boundary is usually enforced via zeroth-order extrapolation, simply copying the state of the innermost active cell into the adjacent ghost cell. This approach implicitly maintains a zero gradient orthogonal to the boundary face. On a mapped grid, however, copying values along the logical grid lines maintains the zero gradient along the direction of the local grid deformation rather than along the true physical face normal which leads to an inaccurate solution as illustrated in figure~\ref{fig:lax_deform}d.

    One potential workaround is to embed the physical domain within a buffer zone of strictly Cartesian cells, ensuring the outer ghost cells remain orthogonal. However, this approach is computationally inefficient and geometrically restrictive.

\begin{figure}
    \centering
    \includegraphics[width=0.5\linewidth]{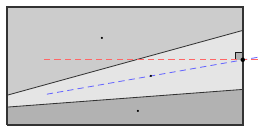}
    \caption{Correction of the outflow boundary conditions on the deformed Cartesian grid. Without modification, the standard ghost cell approach enforces a zero gradient along the deformed logical grid line (blue). Our corrected scheme interpolates the face-centered state using neighboring cells to directly evaluate and impose the boundary flux (red).}
    \label{fig:laxliu_interpolation}
    \end{figure}

    Instead, we address this issue by directly applying the appropriate boundary fluxes rather than relying on ghost cell state extrapolation. We interpolate the primitive state at the exact center of the boundary face using the neighboring cells (Fig.~\ref{fig:laxliu_interpolation}). From this interpolated state, we explicitly compute the absorbing flux through the boundary and recover a more accurate solution even for very deformed grids, as can be readily seen by comparing the solution that includes the correction and displays straighter wave fronts in figure~\ref{fig:lax_deform}e to the reference solution shown in figure~\ref{fig:lax_deform}c. To further demonstrate the quality of the obtained solution on non-Cartesian mappings, we provide in the appendix~\ref{appendix:blast} an additional convergence test on the Sedov blast wave (see figure~\ref{fig:convergence_blast}).
    
\subsection{Rayleigh-Taylor instability}

    The Rayleigh-Taylor instability test is designed to validate the solver's behavior when subjected to gravity-driven hydrodynamic instabilities. The domain is initialized with a high-density fluid resting on top of a lower-density fluid. The system is set in a background hydrostatic equilibrium, onto which a small vertical velocity perturbation is applied at the interface between the heavy and the light fluids. Under the influence of gravity, this configuration is fundamentally unstable: the heavy fluid collapses into the lighter one, forming characteristic plumes, while the light fluid rises.

    \begin{figure}
    \centering
        \begin{subfigure}{0.45\linewidth}
        \includegraphics[width=1\linewidth]{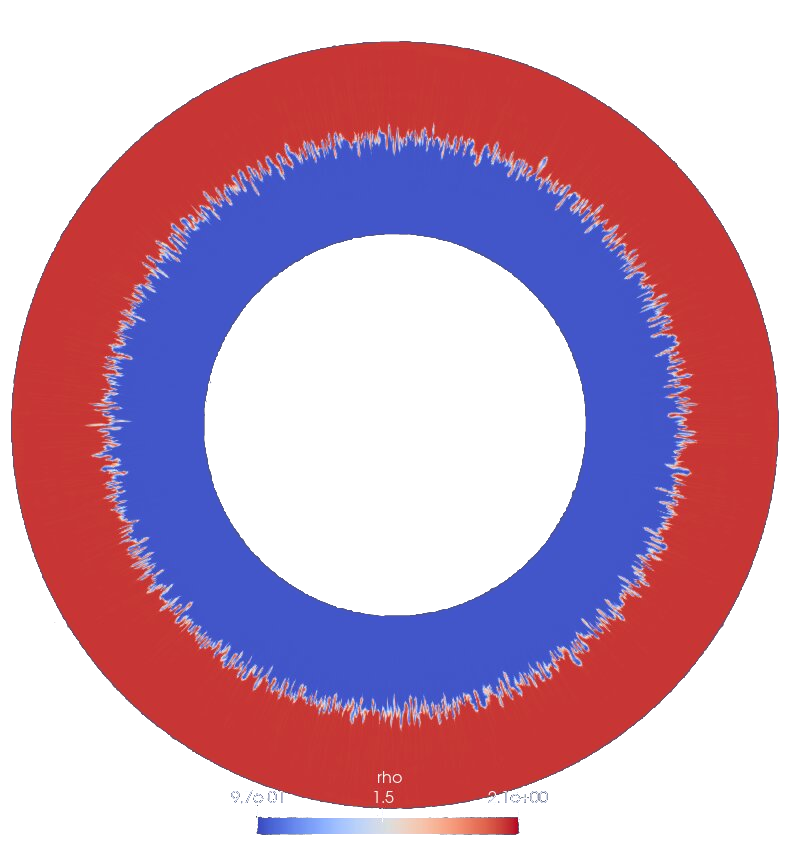}
        \caption{$t = 2\,\mathrm{s}$}
        \end{subfigure}
        \begin{subfigure}{0.45\linewidth}
        \includegraphics[width=1\linewidth]{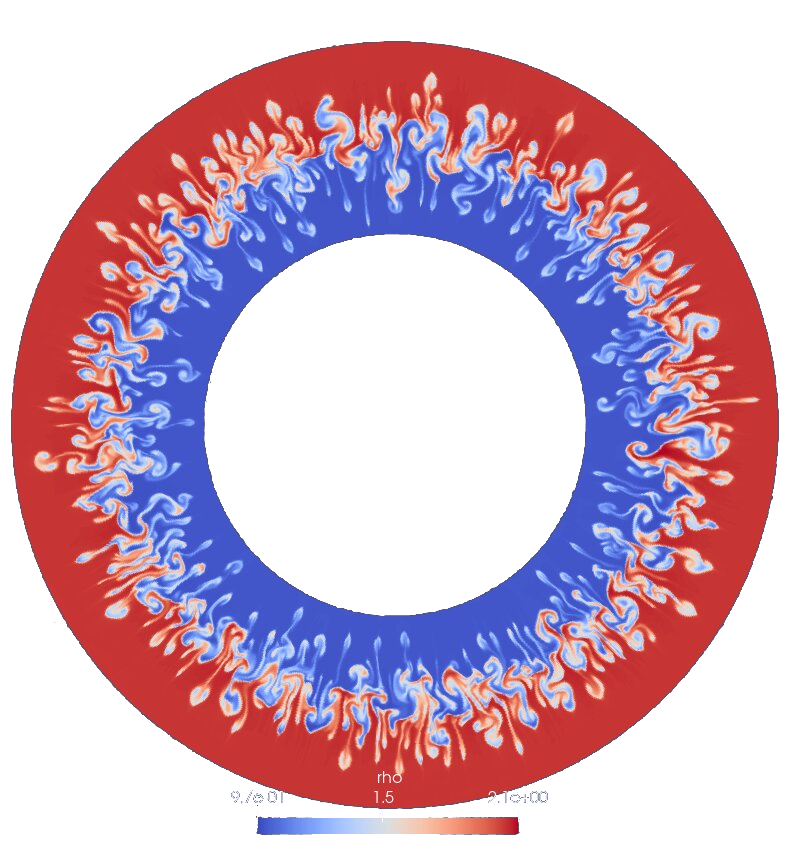}
        \caption{$t = 4\,\mathrm{s}$}
        \end{subfigure}
        \caption{Evolution of the Rayleigh-Taylor instability on a ring grid ($2048 \times 256$, with $r \in [0.5, 1]$). The domain is initialized with a heavy fluid ($\rho_\text{out} = 2$) resting on top of a lighter fluid ($\rho_\text{in} = 1$) under a constant gravity $g = 2/3$. The outer pressure is fixed to $p_\text{out} = 2$, while the internal pressure profile is derived from hydrostatic equilibrium, getting $p_\mathrm{in} = 2.5$. 
        Visualization of the density for $\rho$ between $0.97$ and $2.1$.
        (a) At $t = 2\,\mathrm{s}$, small structures, radially aligned, begin to appear at the contact interface. (b) By $t = 4\,\mathrm{s}$, these structures grow, merge, and lose their radial alignment.}
        \label{fig:rt_ring_evol}
    \end{figure}
    
\begin{figure}
        \centering
        \begin{subfigure}{0.45\linewidth}
        \includegraphics[width=1\linewidth]{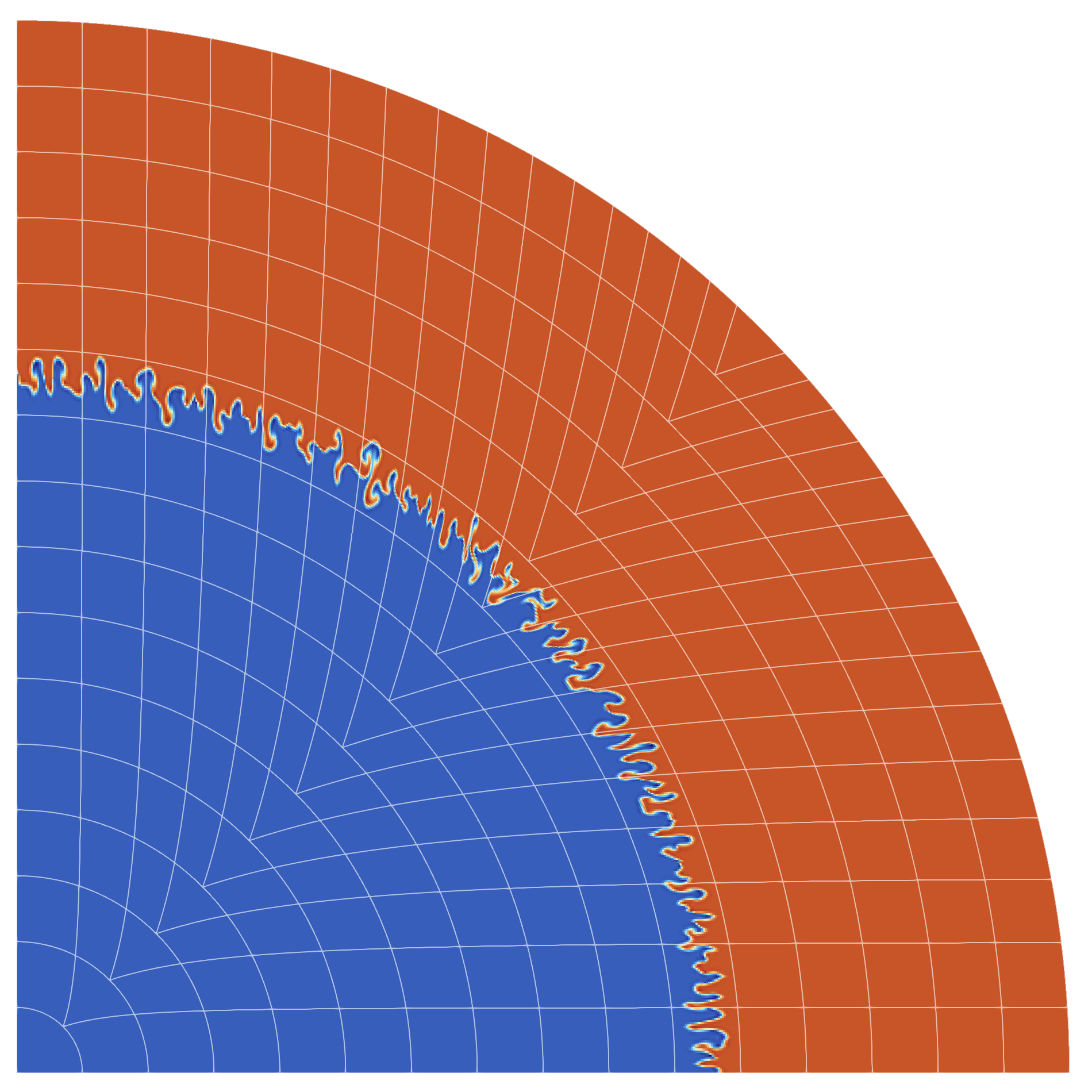}
        \caption{$t = 2\,\mathrm{s}$}
        \label{fig:rt_grid_perturb_a}
        \end{subfigure}
        \begin{subfigure}{0.45\linewidth}
        \includegraphics[width=1\linewidth]{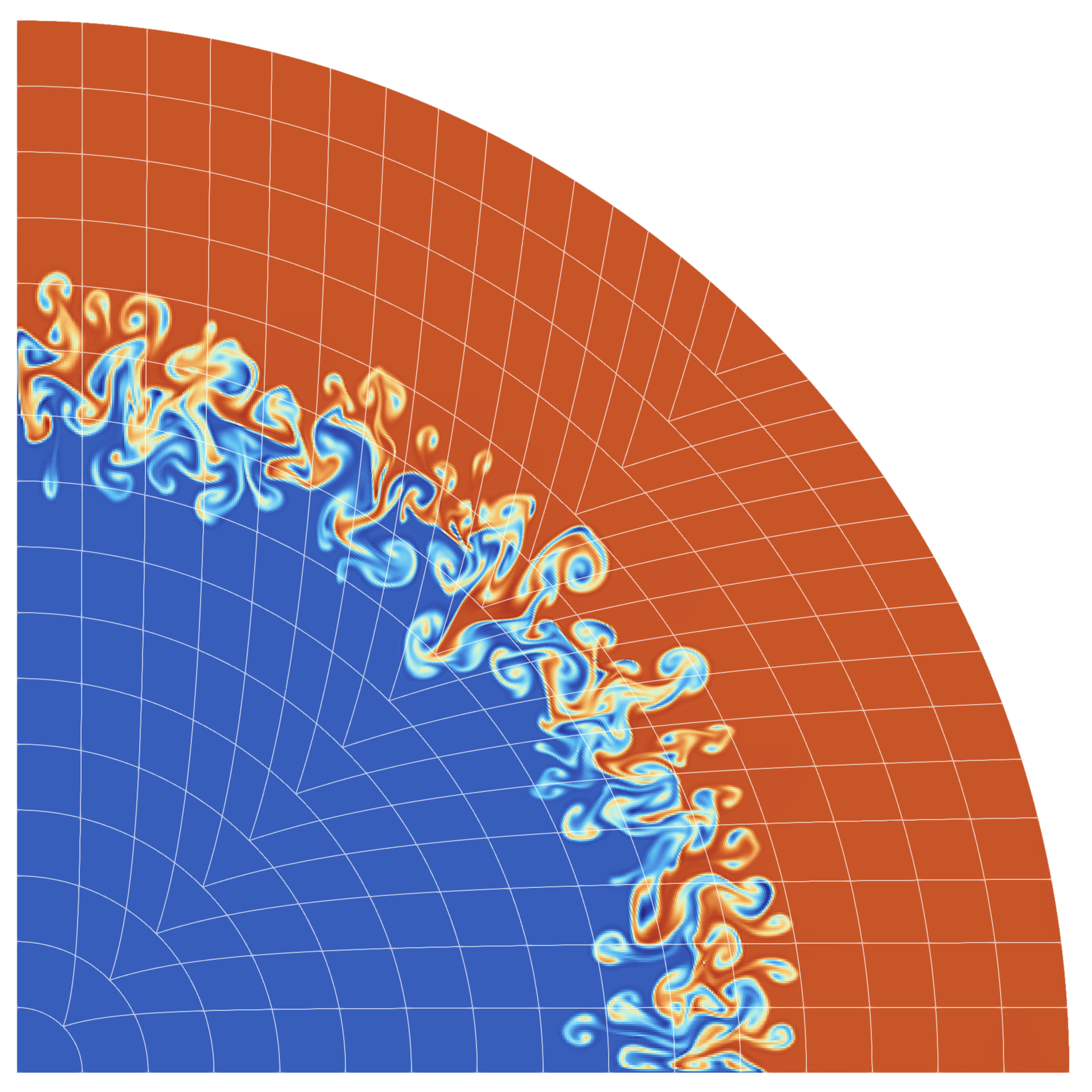}
        \caption{$t = 4\,\mathrm{s}$}
        \label{fig:rt_grid_perturb_b}
        \end{subfigure}
        \begin{subfigure}{0.45\linewidth}
        \includegraphics[width=1\linewidth]{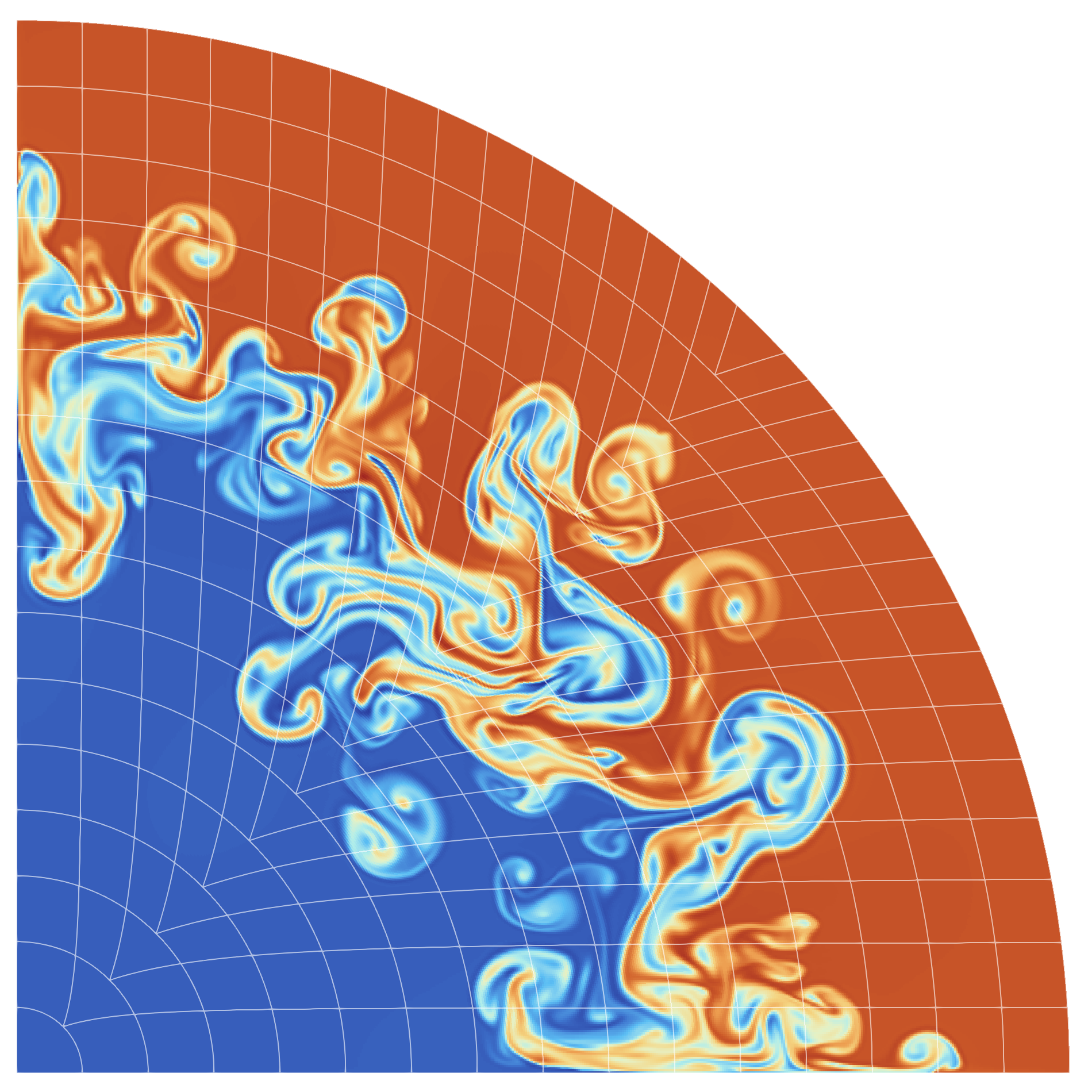}
        \caption{$t = 7\,\mathrm{s}$}
        \label{fig:rt_grid_perturb_c}
        \end{subfigure}
        \begin{subfigure}{0.45\linewidth}
        \includegraphics[width=1\linewidth]{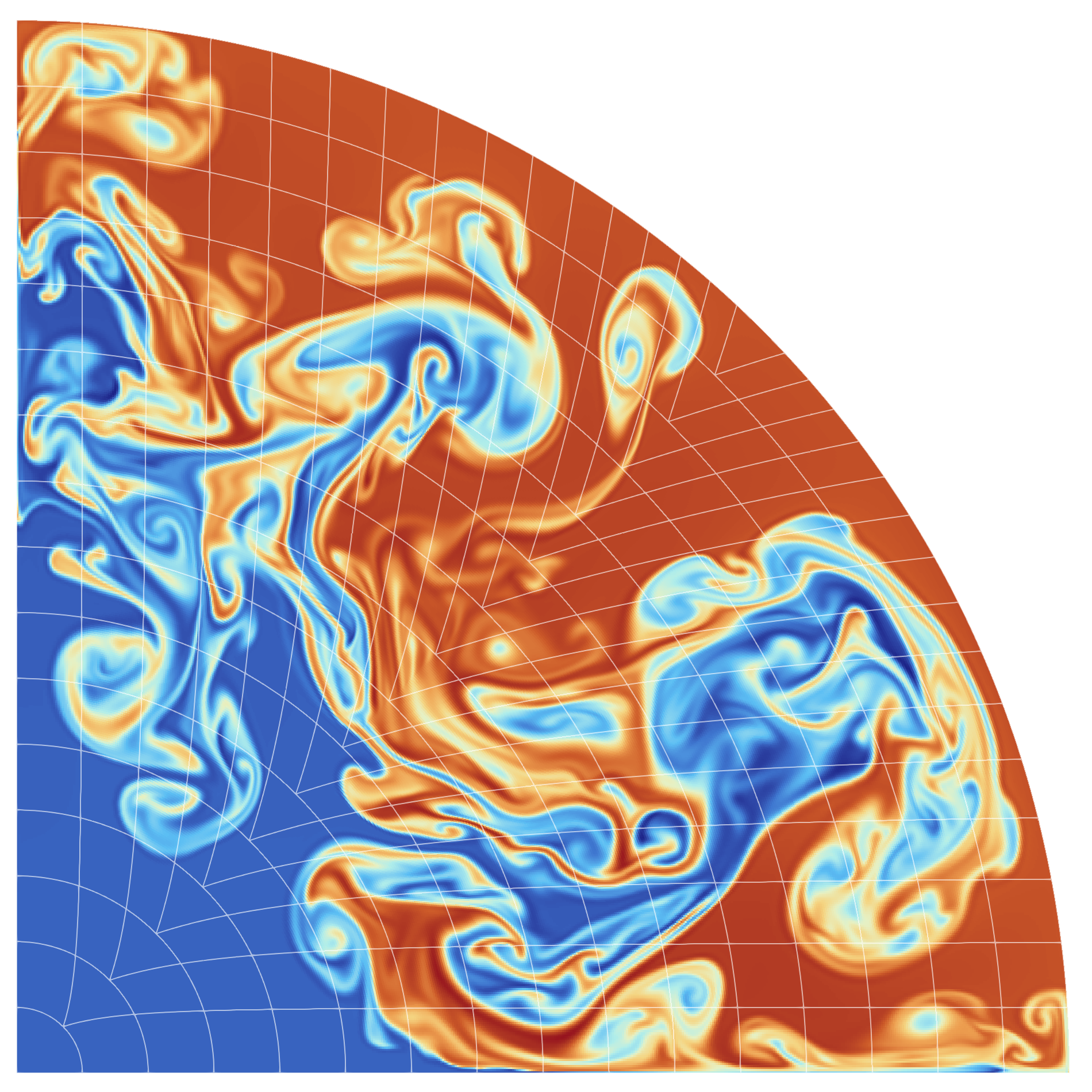}
        \caption{$t = 10\,\mathrm{s}$}
        \label{fig:rt_grid_perturb_d}
        \end{subfigure}
        \caption{Rayleigh--Taylor on the \calhoungrid{} with $r \in \left[0,1\right]$. 
        The gravity is linear of the form $g(r) = 0.5 \,r$. The initial densities are $\rho_\mathrm{in} = 1$, $\rho_\mathrm{out} = 2$. The initial pressure at the surface is $p_\mathrm{out} = 2$, and the pressure at the center is obtained from the hydrostatic equilibrium ($p_\mathrm{in} \approx 2.5$). At $t=2\,\mathrm{s}$, because of the grid deformation, the initial structures tend to be aligned with the grid. For $t\geq4\,\mathrm{s}$, the structures have developed and are less subject to the grid.
        }
    \end{figure}

    The ring grid has the advantage of featuring perfectly regular cells that are exactly aligned with the radial and ortho-radial directions (Fig.~\ref{fig:rt_ring_evol}). However, when we perform this same test on a curvilinear grid such as the \calhoungrid{}, we observe that during the linear growth phase, the initial instabilities tend to propagate by aligning with the preferential directions of the grid's deformation (see Fig.~\ref{fig:rt_grid_perturb_a}).

    Nevertheless, once the instabilities reach the non-linear regime and the size of the structures exceeds the grid scale, turbulent mixing occurs isotropically. The flow fully develops as expected, becoming for $t\geq4\,\mathrm{s}$ less subject to the grid's influence, as can be clearly seen in Fig.~\ref{fig:rt_grid_perturb_c} and ~\ref{fig:rt_grid_perturb_d}.

\subsection{Polytropic convection on a ring grid}
    Before moving toward full global solar simulations, we adapt the classical convection setup of \cite{Cattaneo1991} to a two-dimensional ring domain. This test allows us to extend our validation from pure hydrodynamics to the Navier-Stokes framework by incorporating both thermal conduction and viscosity.
    
    Strict boundary conditions are imposed at both the inner and outer radial edges. For the velocity field, we apply impenetrable, stress-free conditions:
    \begin{equation}
        v_r = 0, \qquad \text{and} \qquad \dv{}{r} \left(\frac{v_\theta}{r}\right) = 0
    \end{equation}
    And for the thermal conduction, we impose Neumann's conditions by fixing the temperature gradient at both boundaries ($\dv{T}{r} = \mathrm{const}$). 

    The background stratification within this bounded domain is initialized in hydrostatic equilibrium as a polytrope:
    \begin{equation}
        T(r) = T_0\, (1 + \theta r), \;
        \rho(r) = \rho_0\, (1 + \theta r)^m, \;
        p(r) = p_0\, (1 + \theta r)^{m+1}
    \end{equation}
    with $m = 1$ and $\theta = 20$, where $m$ is the polytropic index, $T_0$, $\rho_0$, and $p_0$ are the reference values at the inner boundary, and $\theta$ parametrizes the temperature gradient across the layer. To balance this profile, a uniform radial gravitational acceleration is imposed as $\bm{g} = -g_0 \bm{e}_r$, with $g_0 = (m+1) \theta R T_0$.
    
    \begin{figure}
        \centering
        \includegraphics[width=1\linewidth]{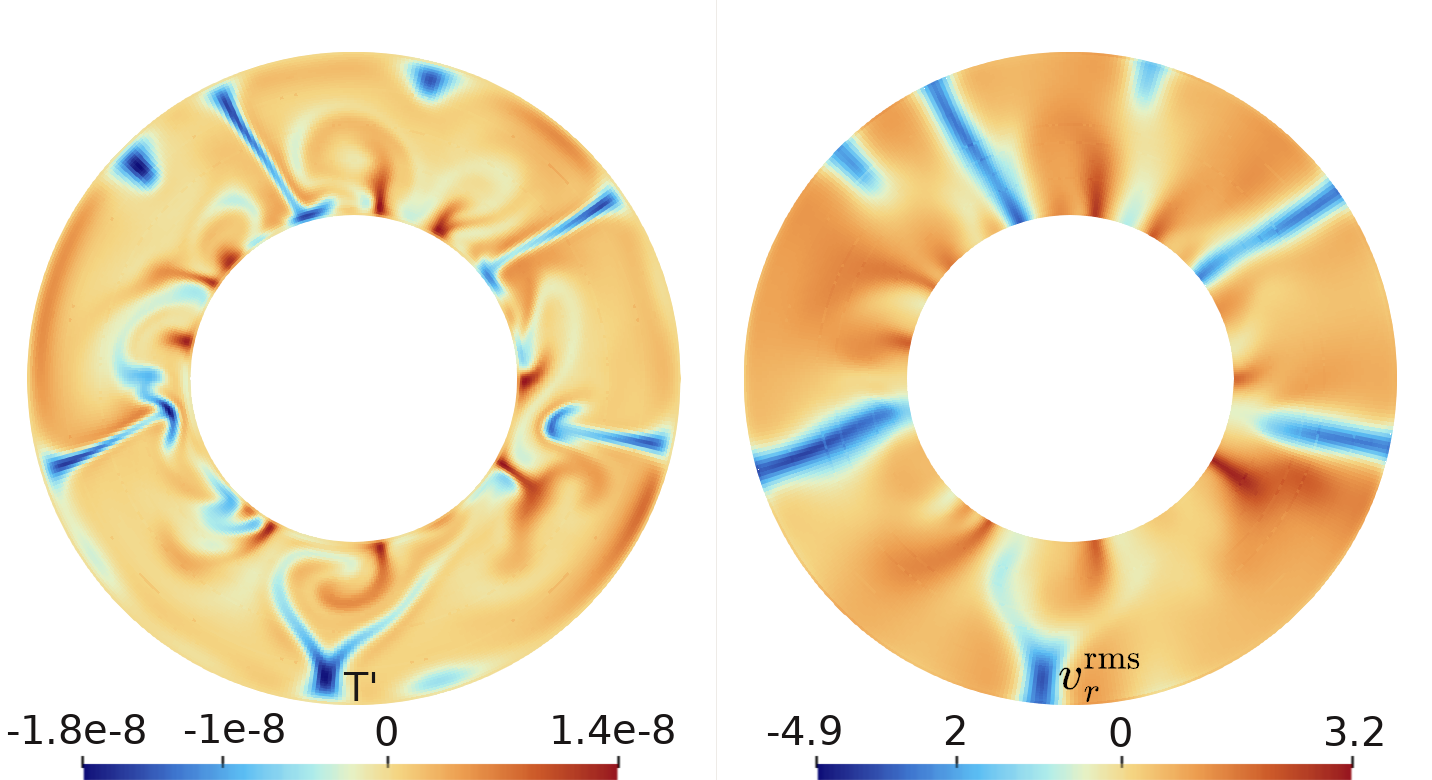}
        \caption{Compressible convection benchmark based on \cite{Cattaneo1991} adapted to a ring grid with parameters $m = 1$ and $\theta = 20$. \textit{Left}: Temperature fluctuations showing the characteristic broad warm upflows and narrow, intense cold downflows. \textit{Right}: Radial velocity.
        }
        \label{fig:c91_ring}
    \end{figure}
    
    The simulation (Fig.~\ref{fig:c91_ring}) successfully develops the hallmark features of compressible convection: narrow, fast downward-flowing cold plumes surrounded by broad, slower upward-flowing warm plumes. The plumes develop an anchor-like shape when splashing into the bottom boundary, triggering return fluxes. These results match the expected dynamics obtained on standard Cartesian grids, thereby validating our metric transformations, gravity source terms, and thermal conduction modules in a curvilinear geometry.
    
    However, this specific shell-based setup inherently relies on the existence of an inner radial boundary. It is not adapted to the $r=0$ case, as a constant radial gravity field is singular at the center. Moreover, polytropic profiles are not the most adapted for solar simulation.
    
    To perform full global simulations using the \calhoungrid{} mapping (Fig.~\ref{fig:map_calhoun}), we require a background convective state that is valid all the way down to the center. The methodology developed to generate such global solar profiles is detailed in the following section.

\subsection{Global solar simulation}
    In this section, we present our coupled convective zone (CZ) and radiative zone (RZ) global compressible convection solar model.

\subsubsection{Initialization setup}
\begin{figure}
        \centering
        \includegraphics[width=0.9\linewidth]{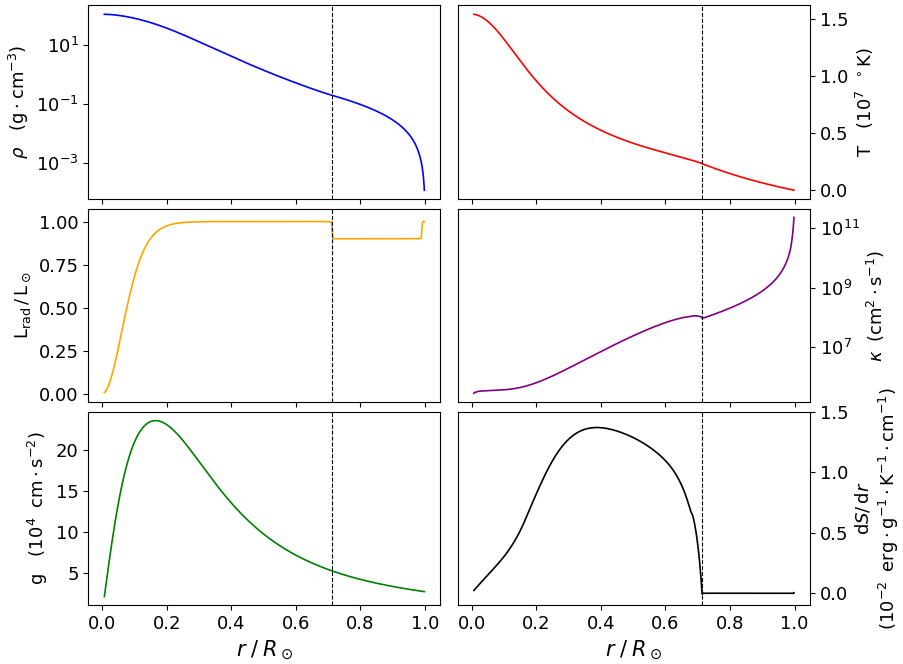}
        \caption{Generated radial profiles for the background state of density, temperature, normalized radiative luminosity, thermal diffusivity $\kappa = K \,/ \left(\rho\,c_p\right)$, gravity, and entropy gradient. 
        The black dashed line marks the limit between the radiative zone and the convection zone (tachocline). }
        \label{fig:solar_profiles}
    \end{figure}

\paragraph{Background state: }
    \label{sec:backgroundstate}
    To initialize our global solar convection simulations, we require a robust method to generate an analytical, hydrostatically and thermally balanced background state that extends from the solar core to the surface. Rather than solving the self-gravitating Lane-Emden equation, we construct our thermodynamic profiles based on prescribed radial profiles of gravity $g(r)$ and entropy gradient $\dd{S} \,/ \dd{r}$ (Fig.~\ref{fig:solar_profiles}). These input profiles are derived from a fully integrated, seismically calibrated Standard Solar Model \citep{Brun2002Antia}.
    Please note that the degree of the super-adiabadicity in the convection zone is still subject of debate as discussed in detail in \cite{2024A&A...682A..39B}.
    
    We assume the solar plasma behaves as an ideal gas with a constant adiabatic index $\gamma = 5/3$ for the whole domain. The background state is governed by the equation of state for entropy $S$, the hydrostatic equilibrium, and the ideal gas relations:
    \begin{equation}
        S = c_v \ln{p} - c_p \ln{\rho}, \qquad
        \dv{p}{r} = -\rho g
    \end{equation}
    where $c_p$ and $c_v$ are respectively the specific heats at constant pressure and volume.
    
    By differentiating the entropy equation with respect to $r$ and substituting the hydrostatic pressure gradient, the system can be combined into a single first-order non-linear ordinary differential equation (ODE) for the density $\rho$ \citep{Miesch2000}:
    \begin{equation}
        \label{eq:background_miesch}
        \dv{\rho}{r} + \frac{\rho}{c_p} \dv{S}{r} + \rho^{2-\gamma}\, \frac{g}{\gamma}e^{-\gamma \,S / c_p} = 0
    \end{equation}

    We solve this equation to obtain our solar background states, and provide the detail of its resolution in the appendix~\ref{anx:solve_background}.
    
    \paragraph{Thermal balance: }
    In addition to the hydrostatic equilibrium, the background state must be in thermal equilibrium. The total energy equation with thermal conduction and a volumetric heating source $q$ is:
    \begin{equation}
        \pdv{E}{t} - \div{(K\grad{T})} = q
    \end{equation}

    Following standard practices in stellar modeling \cite{Miesch2000,Brun2011}, the nuclear energy generation in the core is parameterized as a density- and temperature-dependent power law:
    \begin{equation}
        q(r) = \varepsilon_0 \, \rho\,  T^n
    \end{equation}
    where the constants $\varepsilon_0$ and $n$ are determined by fitting the resulting integrated luminosity to the reference solar luminosity profile.

    To ensure thermal equilibrium, the thermal conductivity profile $K(r)$ must be computed so that the outward radiative flux perfectly transports the generated heat within the radiative interior (Fig.~\ref{fig:solar_profiles}). The total luminosity $L(R)$ generated within a radius $R$ is obtained by integrating the heating term. We must distinguish between the 3D spherical case and the 2D polar case to maintain dimensional consistency in the code:
    \begin{equation}
        L_\text{3D}(R) = 4\pi \int_0^R q(r) r^2 \dd{r}, \quad
        L_\text{2D}(R) = 2\pi \int_0^R q(r) r \dd{r}
    \end{equation}
    
    To trigger convective instability in the outer envelope, we smoothly decrease the radiative luminosity $L_{\text{rad}}$ by a fraction (e.g., 10\%) in the convective zone using a hyperbolic tangent transition (Fig.~\ref{fig:solar_profiles}). In mixing length theory (MLT) this decrease could be as high as 100\%, but here we follow the approach of \cite{Noraz2025} to allow for a better control of the buoyancy in this numerical study.
    
    Finally, the dynamic viscosity profile $\mu(r)$ is derived directly from the thermal conductivity by enforcing a constant Prandtl number ($Pr$) through the whole domain:
    \begin{equation}
        \mu(r) = Pr\, \frac{K(r)}{c_p}
    \end{equation}
    
    \subsubsection{2D global solar convection results}
    All simulations share the same underlying background stratification and setup geometry. The base of the convection zone (tachocline) is located at $r = 0.713\, R_\odot$ \citep{ChristensenDalsgaard1991}, which encompasses roughly $N_\rho = 7$ density scale heights within the convective envelope corresponding to a surface radius $r_s = 0.99\, R_\odot$. The velocity field is subject to stress-free boundary conditions, ensuring strictly zero mass and momentum fluxes across the boundaries, while thermal conduction is closed by  fixing a temperature gradient at the surface. Convection is initiated by introducing a small relative pressure perturbation of $5 \times 10^{-7}$ in the convection zone.
    
\begin{figure}
        \centering
        \includegraphics[width=0.7\linewidth]{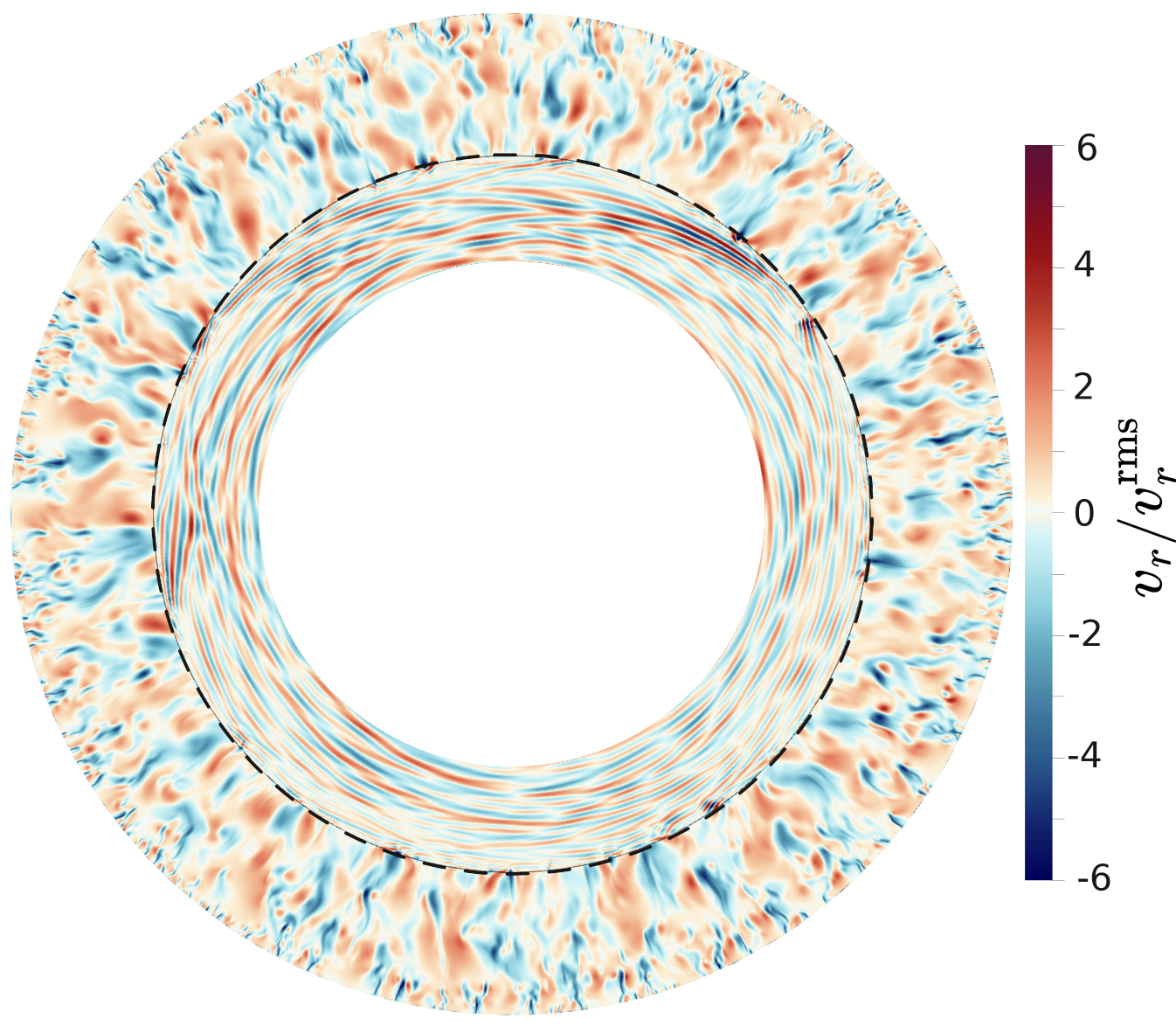}
        \caption{2D solar convection on the uniform ring grid ($4096 \times 512$) with $\rm{Pr}=0.1$. Visualization of the normalized radial velocity, blue for downflows, red for upflows. The dashed line marks the limit between the radiative zone and the convection zone (tachocline). }
        \label{fig:sun_ring}
    \end{figure}

\begin{figure}
        \centering
        \includegraphics[width=0.7\linewidth]{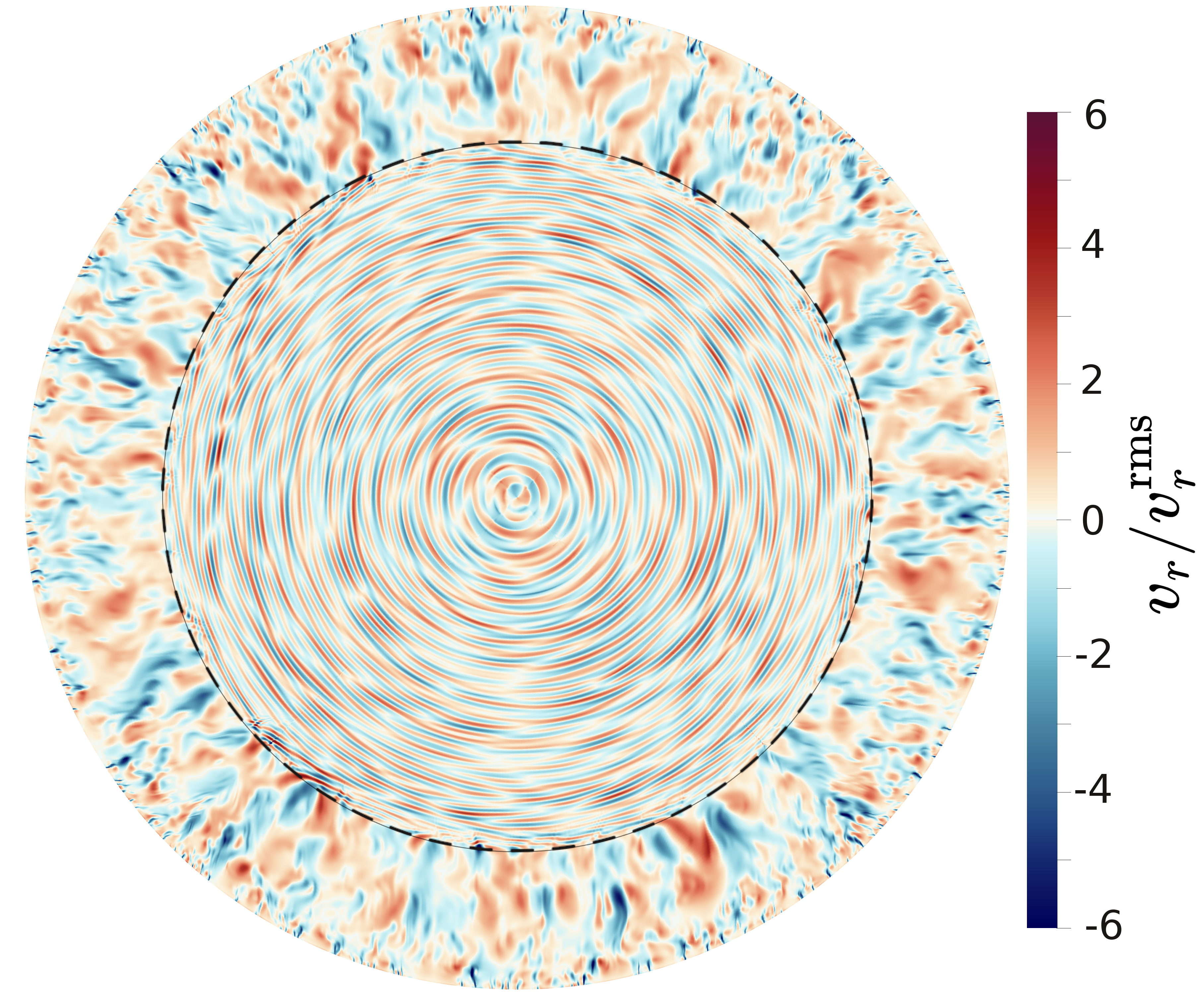}
        \caption{2D solar convection snapshot on the uniform \calhoungrid{} ($2048^2$) with $\rm{Pr}=0.1$. Visualization of the normalized radial velocity, blue for downflows, red for upflows. The dashed line marks the limit between the radiative zone and the convection zone (tachocline).}
        \label{fig:sun_calhoun_uniform}
    \end{figure}

\begin{figure}
        \centering
        \includegraphics[width=0.9\linewidth]{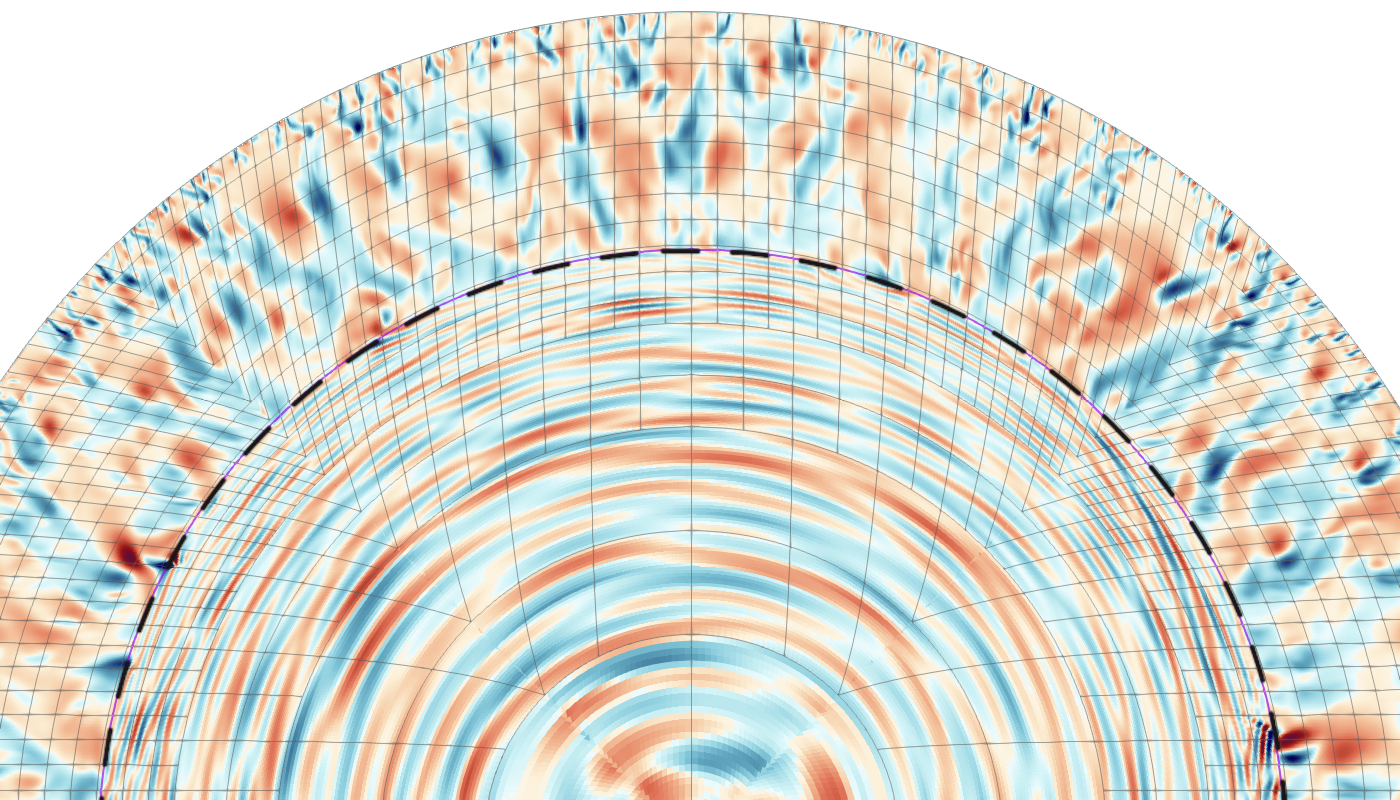}
        \caption{Zoom-in view of a 2D solar convection on the \calhoungrid{} with $\rm{Pr}=0.1$ and 4 levels of static refinement, achieving an effective resolution of $2048^2$ in the convective layer. Each block is made of $32^2$ cells. Visualization of the normalized radial velocity, blue for downflows, red for upflows. The dashed line marks the limit between the radiative zone and the convection zone (tachocline).}
        \label{fig:sun_calhoun_refine}
    \end{figure}

\begin{figure}
        \centering
        \includegraphics[width=0.9\linewidth]{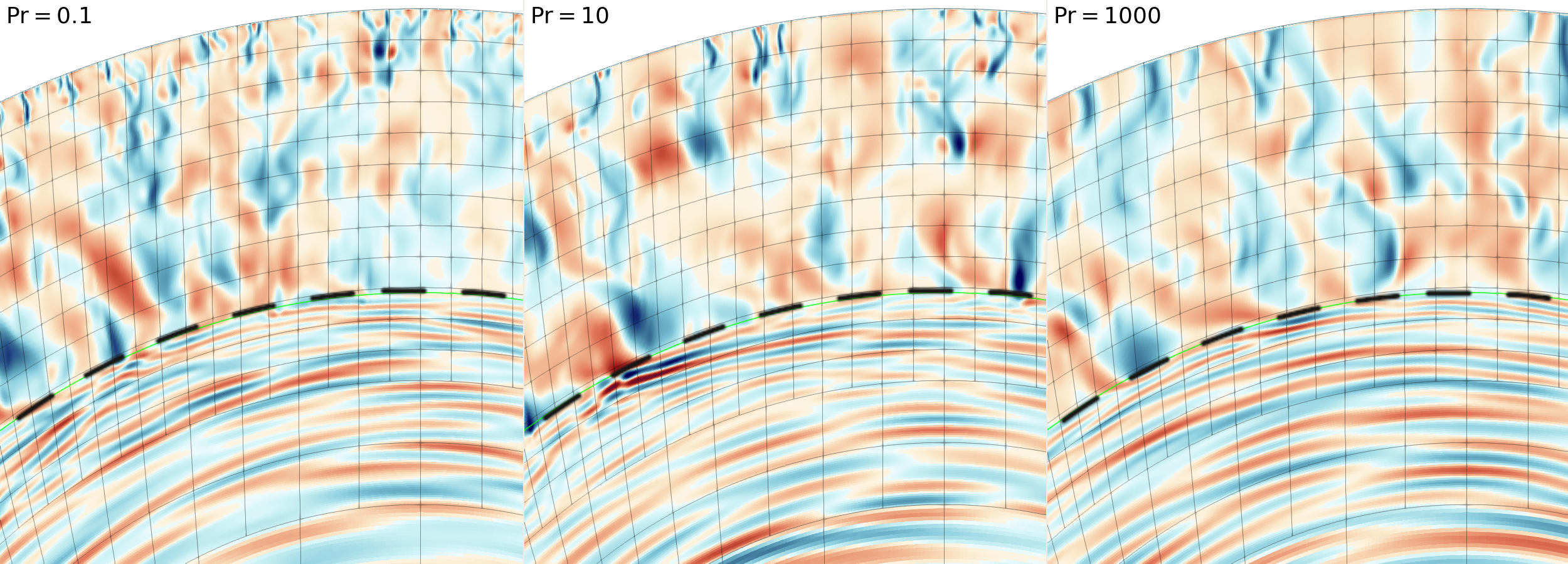}
        \caption{Comparison of different Prandtl number for the solar setup ($\rm{Pr}=0.1$, $\rm{Pr}=10$ and $\rm{Pr}=1000$). 
        The bigger the Prandtl number is, the broader are the plumes. 
        The filling factor of the downward plumes at the surface is of 35\% for the $\rm{Pr}=0.1$ and 30\% for the $\rm{Pr}=1000$, then become the same at 45\% for all $\rm{Pr}$ below $0.9 R_\odot$.
        The behavior in the RZ is similar.
        The dashed line marks the position of the tachocline.
        }
        
        \label{fig:sun_calhoun_refine_multi_prandtl}
    \end{figure}

    To systematically investigate the effect of grid geometry and physical parameters, we compare two baseline uniform configurations against a suite of multi-level statically refined configurations:
    \begin{itemize}
        \item Uniform reference runs: We launch simulations on a regular uniform \textit{ring} grid (resolution $4096 \times 512$) and on a uniform \calhoungrid{} (resolution $2048^2$). Both configurations are executed at a standard low Prandtl number $\rm{Pr} = 0.1$.
        (Fig.~\ref{fig:sun_ring}; Fig.~\ref{fig:sun_calhoun_uniform})
        \item Statically refined runs: Utilizing the AMR capability of \texttt{DWS}, we implement 4 levels of static radial refinement, reaching an effective resolution of $2048^2$ at the surface. On this advanced mesh, we explore distinct Prandtl regimes: $\rm{Pr} = 0.1$, $\rm{Pr} = 10$ and $\rm{Pr} = 1000$.
        (Fig.~\ref{fig:sun_calhoun_refine}; Fig.~\ref{fig:sun_calhoun_refine_multi_prandtl})
    \end{itemize}
    Across all setups, the non-linear convective dynamics are successfully established without triggering any grid-alignment or diagonal-driven numerical artifacts, validating the excellent geometric stability of our well-balanced finite-volume solvers.

    The two uniform grid simulations executed on the ring and \calhoungrid{} (Fig.~\ref{fig:sun_ring}; Fig.~\ref{fig:sun_calhoun_uniform}) exhibit similar convective dynamics. The surface downflows are of comparable side and spread at a similar rate as they descend through the CZ. Furthermore, their interaction with the tachocline is the same, exciting analogous spiral-like internal gravity wave patterns within the stable interior. This excellent agreement demonstrates that \texttt{DWS} successfully reproduces identical solar convective features regardless of the underlying mesh topology, showing high fidelity on both the highly regular ring grid and the significantly less regular \calhoungrid{}.

    In these simulations, the limiting timestep is the hydrodynamic one, as the spatial resolution is not high enough for the $\Delta x^2$ factor of the diffusive timesteps to dominate. The global timestep is therefore constrained at the stellar center, where the sound speed reaches its peak. Consequently, coarsening the grid within the deep radiative zone is highly effective to increase the simulation timestep. The four levels of coarser blocks in the RZ (Fig.~\ref{fig:sun_calhoun_refine}) allowed us to increase the global time step by a factor of approximately five, while maintaining the same high resolution at the surface. With this setup, the limiting hydrodynamic timestep is shifted slightly up within the RZ.
    
    A direct comparison between the uniform and the statically refined setup (Figs.~\ref{fig:sun_calhoun_uniform} and \ref{fig:sun_calhoun_refine}) shows identical fluid dynamics within the CZ and the upper RZ, as expected from their identical grid resolution. Further down in the RZ, where the resolution is reduced compared to the uniform case, the spiral-like shape of gravity waves are consequently broader, yet the overall wave propagation and dynamics remain highly consistent.
    
\begin{figure}
        \centering
        \includegraphics[width=0.9\linewidth]{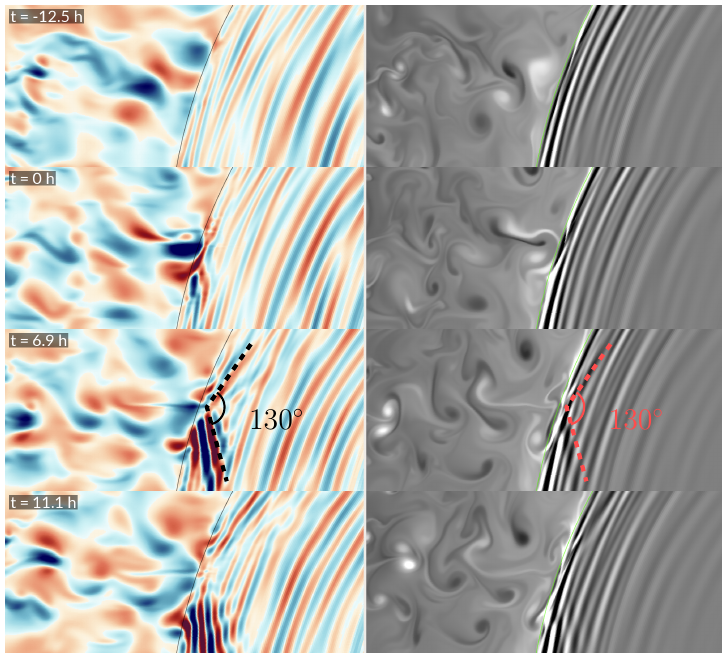}
        \caption{Time sequence of snapshots extracted from the statically refined \calhoungrid{} simulation ($Pr=0.1$). 
        Left: normalized radial velocity (blue downflows, red upflows). Right: temperature fluctuation (white colder, black warmer).
        Highlight a convective plume hitting the tachocline, triggering the formation of a St. Andrew's cross, exciting gravity waves that propagate in the radiative zone following a spiral shape. The spiral propagates in both sides but it is much more stronger in one of them. 
        From top to bottom: (a) downward convective plume. (b) beginning of the interaction with the tachocline. (c) excitation of the gravity waves, forming an angle $\alpha \approx 65^\circ$ between the spiral and the radial direction. (d) post the interaction, propagation of the gravity waves.
        }
    \label{fig:standrew_simulation}
    \end{figure}

    Figure~\ref{fig:sun_calhoun_refine_multi_prandtl} shows that the Prandtl number influences the width of the convective plumes. At low Prandtl numbers, the plumes are much thinner, whereas at high Prandtl numbers, they are significantly broader. At $\rm{Pr} = 0.1$, we observe numerous fine plumes near the surface. At $\rm{Pr} = 10$, the plumes are slightly wider and somewhat less numerous. Finally, at $\rm{Pr} = 1000$, the plumes are clearly broader and fewer.  Conversely, no notable differences in internal wave patterns are observed in the RZ, since the velocity amplitude near the base of the convection zones are very close, exciting internal waves in similar manner (see Fig  \ref{fig:standrew_simulation} for more details on the excitation mechanism). We further noticed that for cases below $\rm{Pr} = 10$, the spectra (see Fig \ref{fig:fftspatial}) tend to converge which is an indication of a numerical floor reached for the viscosity in the simulation for the resolution considered. Therefore, the difference of turbulence degree is likely smaller than the Prandtl numbers difference in the low Pr cases would otherwise indicate.
    
\paragraph{Excitation of gravity waves:}
    As the highly turbulent convection develops in the outer envelope, the flow organizes into fast, narrow, and cold downward-flowing plumes surrounded by broader, slower warm upflows. When these intense downflows reach the base of the convection zone, they overshoot and abruptly impact the stable radiative interior. 
    This perturbation is restored by the buoyancy force, making it act like a periodic oscillator, and effectively exciting internal gravity waves ($g$-modes) that propagate deep into the stellar core \citep{alvan2014, Rogers2006}.

    When the plume interacts with the radiative zone, it forms a cross-like shape known as a \textit{St.~Andrew's cross}. The perturbation propagates in the radiative zone following a spiral. The angle $\alpha$ between the spiral and the radial direction is given by $\alpha = \arccos\left(\frac{\omega}{N}\right)$, where $\omega$ is the wave frequency, and $N$ is the Brunt-Väisälä frequency. 

    Figure~\ref{fig:standrew_simulation} shows consecutive frames of the interaction of a strong downward plume with the radiative zone forming a St.~Andrew's cross. This cross-like perturbation occur over a duration of approximately 10~hours and propagate toward the interior with a clear expansion following a spiral path. The St.~Andrew's cross present a total angle of about $130^\circ$, which correspond to a propagation angle of $\alpha \approx 65^\circ$. In reality, this angle $\alpha$ in the solar interior is close to $90^\circ$, corresponding to low frequency waves ($\omega \ll N$). Therefore, the value of $65^\circ$ obtained here indicates that the plume impact triggers a relatively high-frequency wave regime for the solar interior.
    
\begin{figure}
        \centering
        \includegraphics[width=0.9\linewidth]{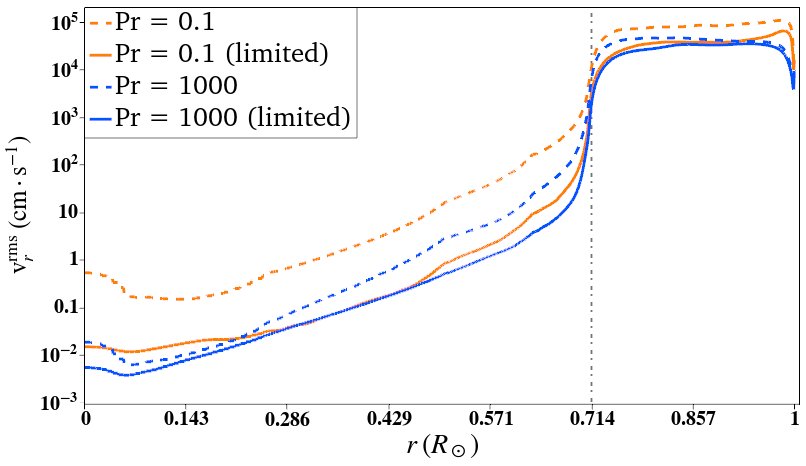}
        \caption{ Time-averaged normalized (RMS) radial velocity $v_r^{\mathrm{rms}}$ of the statically refined model on the \calhoungrid{} for low and high Prandtl numbers ($Pr=0.1$, $Pr=1000$), with and without a velocity limiter enabled in the convective zone.
            }
        \label{fig:radialvelocityrms}
    \end{figure}

   Figure \ref{fig:radialvelocityrms} displays the time-averaged radial profiles of the root-mean-square radial velocity ($v_r^{\mathrm{rms}}$) realized in two representative solar models with different Prandtl numbers, one case with $\mathrm{Pr}=0.1$ in orange and the other case with $\mathrm{Pr}=1000$ in blue. The velocities reach amplitudes between 3 to $5 \times 10^4\;\mathrm{ cm}\cdot s^{-1}$ in the convective envelope. The velocity then drops by several orders of magnitude at the base of the convective envelope $\sim 0.713\, R_\odot$, and by up to six orders of magnitude near the center. The model with $\mathrm{Pr}=1000$ presents a lower and more uniform velocity across the convection zone and the drop in the radiative zone is also more pronounced. We also display the same models after having turned on the velocity limiter. We see that the amplitude of the flows are even lower and the drop stronger, being slightly below the unlimited $\mathrm{Pr}=1000$ case, and getting closer to \citep{alvan2014} whose lower velocity amplitude is likely due to higher viscosity levels and a smaller number of density scale heights in the convective envelope than in the models presented here ($N_\rho$ 3 vs 7).
    
\paragraph{Helioseismology and spectral analysis:}
    A major advantage of our fully compressible approach over traditional anelastic models is the preservation of both gravity waves (buoyancy--driven) and acoustic waves (pressure gradient--driven) without the filtering limitations inherent to anelastic approximations \citep{2012ApJ...756..109B}. To identify the waves modes in a simulation, we record a high frequency timeseries of the radial velocity field $v_r$ at a fixed radius $r$.
    The $\Delta t$ between each snapshot is chosen such that we can recover the high time frequency mode (acoustic mode) from the sampling by following the Nyquist–Shannon theorem, in this case we used $\Delta t = 100\,\mathrm{s}$ corresponding to a Nyquist frequency of $1/(2\Delta t) =  5000 \, \mu\rm{Hz}$ for a total timeseries length of $10^6 \, \mathrm{s}$. The following analysis is performed on the high-resolution uniform simulation shown in Fig.~\ref{fig:sun_calhoun_uniform} ($\mathrm{Pr}=0.1$).
    
\begin{figure}
    \centering
    \includegraphics[width=0.9\linewidth]{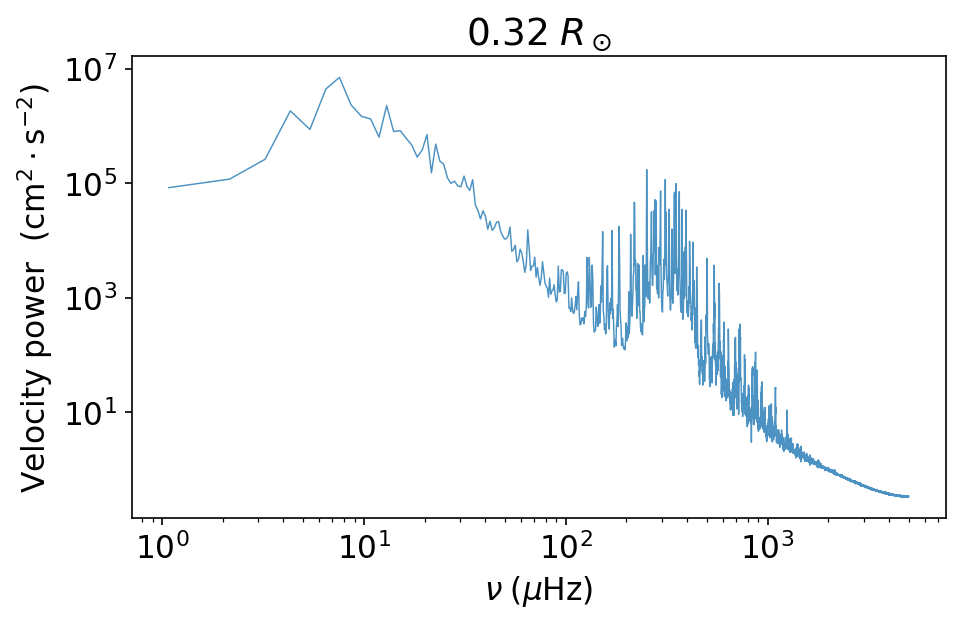}
    \caption{Power spectrum of the radial velocity at $r = 0.32\, R_\odot$. The power excess between $100\,\mu\rm{Hz}$ and $1500\,\mu\rm{Hz}$ characterize the presence of p and g-modes. }
    \label{fig:golf_spectre}
\end{figure}

\begin{figure}
    \centering
    \includegraphics[width=0.9\linewidth]{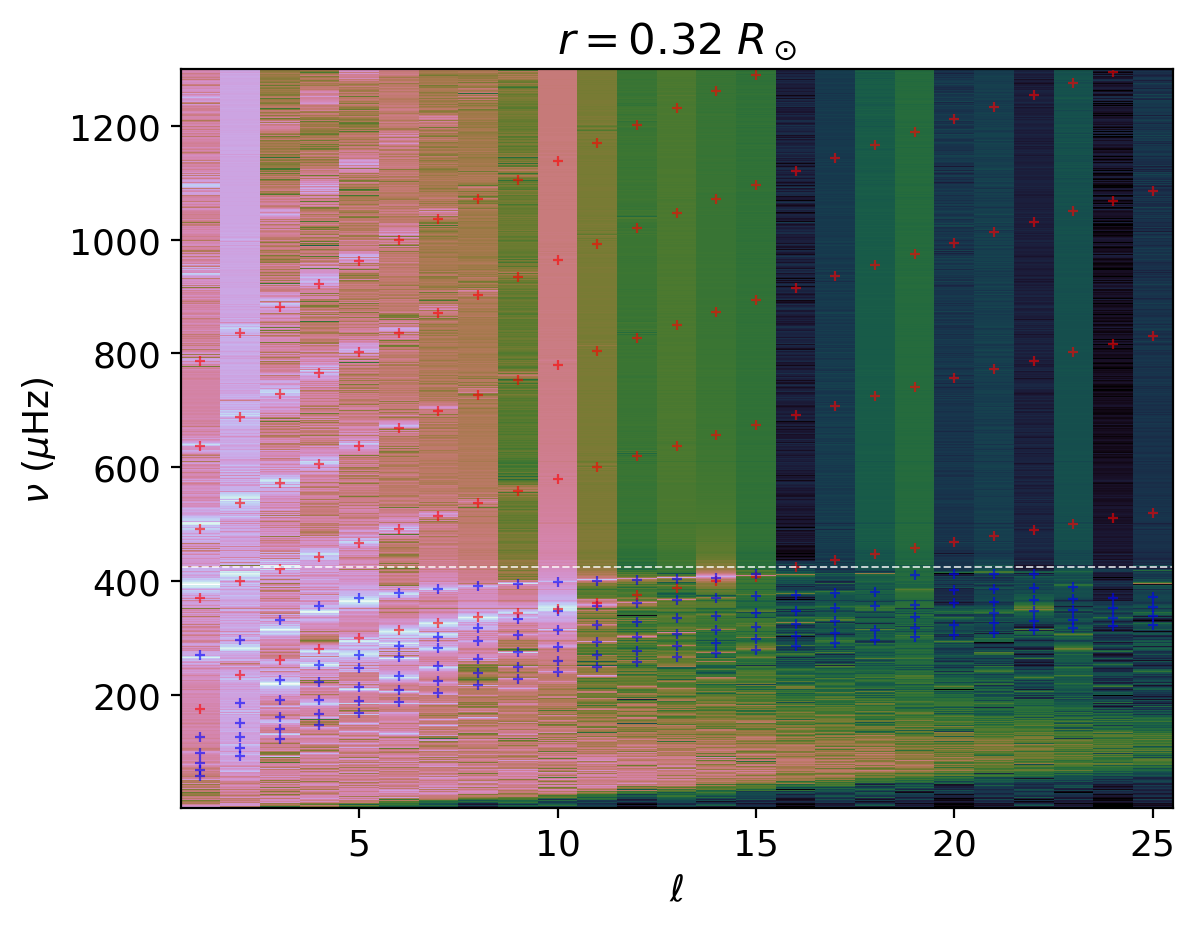}
    \caption{Power spectrum of the simulated solar velocity fluctuations computed at $r=0.32\, R_\odot$ as a function of spatial frequency $\ell$ and waves frequency $\nu$. The background color maps the simulated power spectrum, while the overlaid dots represent the theoretical eigen--frequencies calculated with the GYRE code \citep{Townsend2013}; with $p$ modes in red, and $g$ modes in blue). The white dashed line corresponds to the value of the Brunt-Väisälä frequency at this radius (see Fig.~\ref{fig:brunt_lamb}). The modes frequencies obtained in the simulation and GYRE are in excellent agreement, even though the signal becomes weak at high spatial frequencies.}
    \label{fig:power_spectrum}
\end{figure}

\begin{figure}
    \centering
    \includegraphics[width=0.95\linewidth]{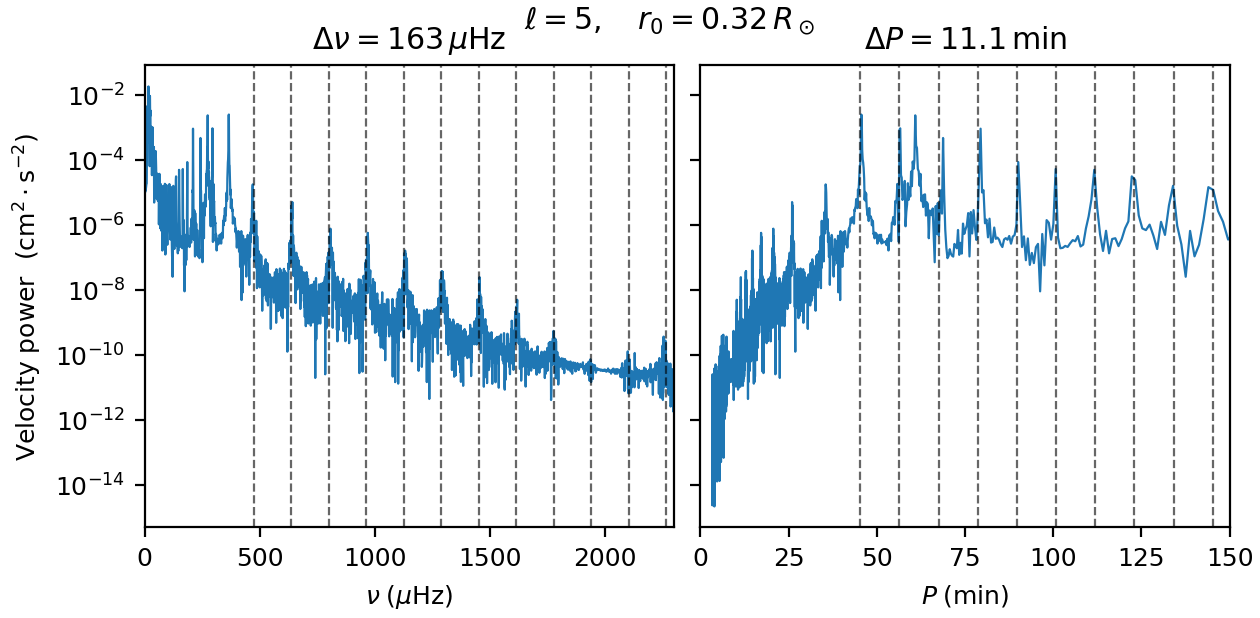}
    \caption{Power spectrum of the radial velocity at $r=0.32 \,R_\odot$ and $\ell = 5$ shown in frequency and in periods. The p-modes are regularly spaced in frequencies, whereas the g-modes are regularly spaced in periods. We see here that the spacing in frequency for the acoustic mode is approximately $\Delta \nu = 163 \, \mu \rm{Hz}$, and the spacing in period for the gravity mode is $\Delta P_5 = 11.1 \, \rm{min}$, in excellent agreement with the theoretical equations (\ref{eq:sismo}). }
    \label{fig:power_spectre_slice}
\end{figure}

    In figure~\ref{fig:golf_spectre}, we show the power spectrum as a function of the degree $\ell$ at $r = 0.32\,R_\odot$. We clearly see an excess power between $100\,\mu\mathrm{Hz}$ and $1500 \,\mu\mathrm{Hz}$ characteristic of the presence of oscillations.
    By calculating the 2D power spectrum at fixed radius $r$, we show $(\ell, \nu)$ diagram \citep{alvan2014} of the simulation in figure~\ref{fig:power_spectrum}. In addition, we use the GYRE stellar oscillation code \citep{Townsend2013} to compute the theoretical eigen-frequencies that are propagating in the domain. We observe an excellent agreement between the modes from the simulation and the theoretical modes obtained from GYRE with obvious ridges corresponding to specific radial order $n$.
    In Fig.~\ref{fig:power_spectre_slice}, we show the computed power spectrum at a radius of $r = 0.32\,R_\odot$ for $\ell=5$, plotted in both frequency and period domains. As expected, the acoustic modes are regularly spaced in frequency, whereas the gravity modes are regularly spaced in period. For this specific setup, we measure a large acoustic separation of $\Delta \nu = 163 \, \mu\rm{Hz}$ and a characteristic period spacing of $\Delta P_5 = 11.1 \, \rm{min}$.
    
    We computed the theoretical values of $\Delta \nu$ and $\Delta P_\ell$ directly from the radial profiles using the following formula \citep{Tassoul1980,Garcia2019}:
    \begin{equation}\label{eq:sismo}
        \Delta \nu = \left[ 2 \int_0^{R} \frac{\dd r}{c_s} \right]^{-1}, \;
        \Delta P_\ell = \frac{\pi}{\sqrt{\ell (\ell+1)} \int_{r_1}^{r_2} \frac{N}{r} \,\dd r}
    \end{equation}
    where $c_s$ is the sound speed. We show that the resonance cavity extends up to $0.99 R_\odot$ as delimited by the acoustic cutoff frequency. For the g-modes, the computed period corresponds to a cavity spanning almost the entire radiative interior.
    
    This constitutes the first joint observation of acoustic modes ($p$--modes) and internal gravity modes ($g$--modes) in a global solar simulation validating the fully compressible nature of the framework and extending on previous global solar convection simulations as in \cite{alvan2014}.

\section{Discussion and conclusion}

We have described the extension of the Cartesian finite-volume framework \texttt{Dyablo} to its global stellar version \texttt{Whole Sun} (\texttt{DWS}) for non-Cartesian geometries using a mapped grid approach to model stars. The standard algorithm was adapted to solve local Riemann problems by rotating the vector quantities into the logical frame of the cell interfaces. We have in particular considered the \calhoungrid{} to model the Sun as a fully global domain extending to the stellar core.

While mapped quadrilateral grids have valuable properties for spherical topologies, the mapping introduces highly skewed cells, particularly along the diagonals. In order to mitigate spurious numerical velocities and retain accuracy in the presence of strong gravitational source terms, we have introduced the $\alpha$--$\beta$ well-balanced scheme perfectly suited for irregular grids and AMR transitions. We provided numerical evidence for the robustness of our scheme through standard test problems, including the 2D Riemann test and the Rayleigh-Taylor instability.

Furthermore, we detailed a mathematical methodology to generate global solar background profiles, and we showed that our method works quite well for simulating compressible turbulent convection in a global 2D solar model. 
Our simulations successfully capture the complex dynamical coupling between the convection zone and the radiative interior, including the overshooting of convective plumes and the excitation of internal gravity waves at the base of the convection zone. 
We have identified that the limiter used in the convection zone influences the amplitude of the velocity, and we intend to explore this aspect in the near future.
One key advantage of performing fully compressible simulations as presented in this study is to retain the acoustic waves that have been observed in the Sun. 
We have shown, that the property of the g and p-modes obtained in the \texttt{DWS} simulation match very well linear eigen-frequencies computed with the GYRE code. In particular the large separation $\Delta \nu$ and gravity period spacing $\Delta P_\ell$ are well reproduced.

While we demonstrated in this paper that the non Cartesian mapping method works well in the hydrodynamic regime to model the Sun, it naturally  extends to magnetized plasmas. The method  presented here is fully compatible with \texttt{Dyablo}'s existing MHD modules, opening the door to global dynamo simulations. We will perform such MHD simulations in the near future. 
Finally, we also plan to conduct detailed helioseismic wave analyses and extend the \texttt{DWS} framework to fully 3D rotating spherical simulations.
\begin{acknowledgements}
We thank GENCI for giving us access to supercomputers at Idris and TGCC centers via the 1623 project. G.D wishes to thank S. Breton for useful discussion and for help in using the open source stellar oscillation code GYRE \citep{Townsend2013}. All authors acknowledges funding by the ERC synergy grant Whole Sun \#810218. All authors thank A. Strugarek, A. Nordlund, M. Carlsson, B. Gudiksen, A. Popovas, M. Szydlarski, R. Cameron, N. Vitas, A. Finley, B.P. Brown, H. Hotta, P. Kestener, N. Aunai, Q. Noraz and L. Barbier for useful discussions and are grateful to both CNES via its Solar Orbiter program and to CNRS INSU-AT-ST for their financial supports. We are thankful to an anonymous referee for their constructive comments that led to an improved version of the paper.
\end{acknowledgements}
\bibliographystyle{aa}
\bibliography{biblio}

\appendix
\nolinenumbers

\section{Dyablo}
\label{anx:dyablo}
\texttt{Dyablo} is a code for the simulation of astrophysical plasmas on exascale architectures using Adaptive Mesh Refinement (AMR). The code targets new and upcoming Exascale supercomputers with a code base in C\texttt{++} using the Kokkos \citep{Trott2022} portability performance library. Kokkos allows us to write a unique code to target multiple hardware architectures including GPUs. Thus, \texttt{Dyablo} is able to run on any modern supercomputer, regardless of vendor or architecture. The design philosophy of \texttt{Dyablo} is based on modern software engineering principles to allow ``separation of concerns'' and collaboration between HPC engineers, physicists and numericists through abstract interfaces and modularity. 

The AMR in \texttt{Dyablo} is block-structured and tree-based. The refinement is operated on octants organized as a linear octree. The linear tree stores only the leaves in the Morton Z-curve order. The block-structure means the leaves of the tree are regular Cartesian blocks. Increasing the size of these regular Cartesian blocks serves two purpose: 1. Increasing the regularity of the mesh allows a better streamlining of operations on the GPU and 2. Having larger blocks allows us to decrease the depth of a tree for a given target resolution. Having a shallower tree helps for load-balancing across multiple processes. The structure and the physical data live on the device so that, in the case of a GPU run, no transfer between CPU and GPU are operated except for IOs. IOs are currently being written to a XDMF/HDF5 unstructured format compatible with most visualization software. Finally, \texttt{Dyablo} comes with a Python analysis package called \texttt{Pyablo} allowing the users to read snapshots, timeseries, and process the data in parallel.

The core of \texttt{Dyablo} groups common \textit{plugins} (ie interchangeable software feature bricks) necessary to most applications. Each dedicated application has their own branch with their own sets of plugins suited to their research and numerical experiments. This paper is centered more specifically around the Whole Sun branch of the code dedicated to solar and stellar physics, named \texttt{Dyablo Whole Sun} or \texttt{DWS}. The base branch of the code is available on github\footnote{https://github.com/Dyablo-HPC/Dyablo}. The code presented in this paper will be made available publicly in coming months.

\section{Geometric properties 3D}
\label{anx:geo3d}

\begin{itemize}
    \item Volume: 
        The volume of a cell is computed as a tetrakis hexahedron as in \cite{Grandy1997}.
        \begin{equation*}
            12 \,V = 16\, v_1 + d_1 + d_2
        \end{equation*}
        where $v_1 = \det(\Delta \bm{f}_x, \Delta \bm{f}_y, \Delta \bm{f}_z)$ with $\Delta \bm{f}_d$ being the vector connecting opposite face centers along direction $d$. The diagonal terms are 
         \begin{equation*}
            \begin{split}
              d_1 &= \det(\bm{r}_5-\bm{r}_0, \bm{r}_7-\bm{r}_0, \bm{r}_2-\bm{r}_0) \\
              d_2 &= \det(\bm{r}_6-\bm{r}_3, \bm{r}_6-\bm{r}_1, \bm{r}_6-\bm{r}_4)
            \end{split}
         \end{equation*}
    \item Face area: 
        The oriented face area (toward the exterior of the cell) is computed with the cross product of its diagonals:
        \begin{equation*}
            \bm{A}_f = \frac{1}{2} (\bm{r}_{f,2} - \bm{r}_{f,0}) \times (\bm{r}_{f,3} - \bm{r}_{f,1})
        \end{equation*}
        The scalar face area is $A = \norm{\bm{A}_f}$.
    \item Face normal:
        \begin{equation*}
            \bm{n}_f = \frac{\bm{A}_f}{\norm{\bm{A}_f}}
        \end{equation*}
        
    \item Centroid: 
        The centroid $\bm{C} = (C_x, C_y, C_z)$ of a cell is computed using the divergence theorem \citep{Nurnberg2013}, and leads to the following formula:
        \begin{equation*}
            C_d = \frac{1}{32\, V} \sum_{f \in \text{faces}} A_{f,d} \sum_{i=0}^{3} (x^f_{i,d} + x^f_{i+1,d})^2
        \end{equation*}
        where $d \in \{x,y,z\}$, $\bm{r}^f_i$ is the $i^\mathrm{th}$ vertex of the face $f$ ordered counter-clockwise (with $\bm{r}_{f,4} \equiv \bm{r}_{f,0}$), and $\bm{A}_f$ is the oriented face area normal.
        
    \item Rotation matrix \citep{Moller1999}:
        We choose to solve all Riemann problems on the X-direction. The rotation matrix that rotate a face normal on the $+\bm{e}_x$ vector is:
        \begin{equation*}
            R = \begin{pmatrix}
            n_x & n_y & n_z \\
            -n_y & n_x + h\, n_z^2 & -h\, n_y n_z \\
            -n_z & -h\, n_y n_z & n_x + h\, n_y^2
            \end{pmatrix}, \quad \text{where } h = \frac{1}{1 + n_x}
        \end{equation*}
\end{itemize}

    \section{Temporal evolution of the system of equations}
    \label{anx:timescheme}
    \subsection{Time integration scheme}
    To ensure numerical stability and achieve second-order accuracy in time, the system is advanced using a Strong Stability Preserving Runge-Kutta scheme (SSP-RK2) \cite{Gottlieb1998, Gottlieb2001}.

    This predictor-corrector scheme is particularly well-suited for hyperbolic conservation laws, as it is constructed by combining forward Euler steps while maintaining the Total Variation Diminishing (TVD) properties of the spatial discretization \citep{Shu1988}. This non-linear stability property guarantees that the time integration preserves strict monotonicity, preventing the generation of spurious numerical oscillations. 
    
    Let $\mathcal{L}(\Uqt)$ denote the spatial discretization operator, which includes the divergence of fluxes and source terms, such that a standard explicit Euler update is written as $\Uqt^{n+1} = \Uqt^n + \Delta t \,\mathcal{L}(\Uqt^n)$.

    The SSP-RK2 scheme consists of two explicit prediction steps followed by a correction step:
    \begin{itemize}
        \item Prediction steps:
        \begin{equation*}
            \begin{split}
                \Uqt^*    &= \Uqt^n + \Delta t \mathcal{L}(\Uqt^n) \\
                \Uqt^{**} &= \Uqt^* + \Delta t \mathcal{L}(\Uqt^*)
            \end{split}
        \end{equation*}
        \item Correction step:
        \begin{equation*}
            \Uqt^{n+1} = \frac{1}{2} \left( \Uqt^n + \Uqt^{**} \right)
        \end{equation*}
    \end{itemize}
    
    \subsection{Time step constraints}
    The global simulation time step is restricted by the Courant-Friedrichs-Lewy (CFL) condition. To ensure numerical stability, this condition limits the time step to prevent physical information from crossing an entire cell within a single iteration. In a standard Cartesian framework, the hydrodynamic time step is typically evaluated component-wise using the cell widths $(\Delta x, \Delta y, \Delta z)$ and the velocity $(u,v,w)$.

    However, in a mapped grid, this approach is less straightforward. The physical cell width does not necessarily represent the shortest characteristic length of a distorted cell (for instance, in highly skewed cells, the diagonal may be shorter). Furthermore, velocities are expressed in the global $(x,y,z)$ reference frame rather than the cell's local topological frame, which would require costly geometric projections.
    
    To bypass these complications, we use an isotropic characteristic cell length $h$ combined with the velocity norm $\norm{\bm{u}}$. The hydrodynamic time step restriction is expressed as:
    \begin{equation}
        \Delta t_\text{hydro} = \mathcal{C}_\text{hydro}\,\frac{h}{c_s + \norm{\bm{u}}}
    \end{equation}
    where $\mathcal{C}_\text{hydro}$ is the CFL number, $c_s = \sqrt{\gamma p / \rho}$ is the sound speed, and $h$ is based on the cell volume $V$ ($h = \sqrt{V}$ in 2D, and $h = \sqrt[3]{V}$ in 3D).
    
    For a uniform grid, $\Delta t_\text{hydro}$ is severely limited at the center of the domain where the high temperature drastically increases $c_s$. We utilized the Adaptive Mesh Refinement (AMR) capability of \texttt{Dyablo} to coarsen the grid in this central region, increasing the characteristic cell length $h$ and the $\Delta t_\text{hydro}$.

    Similarly, the system is also constrained by the parabolic terms from thermal conduction and viscosity:
    \begin{equation}
        \Delta t_\text{th} = \mathcal{C}_\text{th} \, \frac{h^2}{2 \kappa}, \qquad \text{and} \qquad \Delta t_\text{vis} = \mathcal{C}_\text{vis} \, \frac{h^2}{2 \nu}
    \end{equation}
    where $\kappa = K / (\rho C_v)$ is the thermal diffusivity, and $\nu = \mu / \rho$ is the kinematic viscosity. 

    The solar convection simulations presented in this work are performed with a CFL number of $\mathcal{C}_\text{hydro} = \mathcal{C}_\text{th} = \mathcal{C}_\text{vis} = 0.3$.  

    \subsection{Temporal conservation}
\begin{figure}
        \centering
        \includegraphics[width=0.94\linewidth]{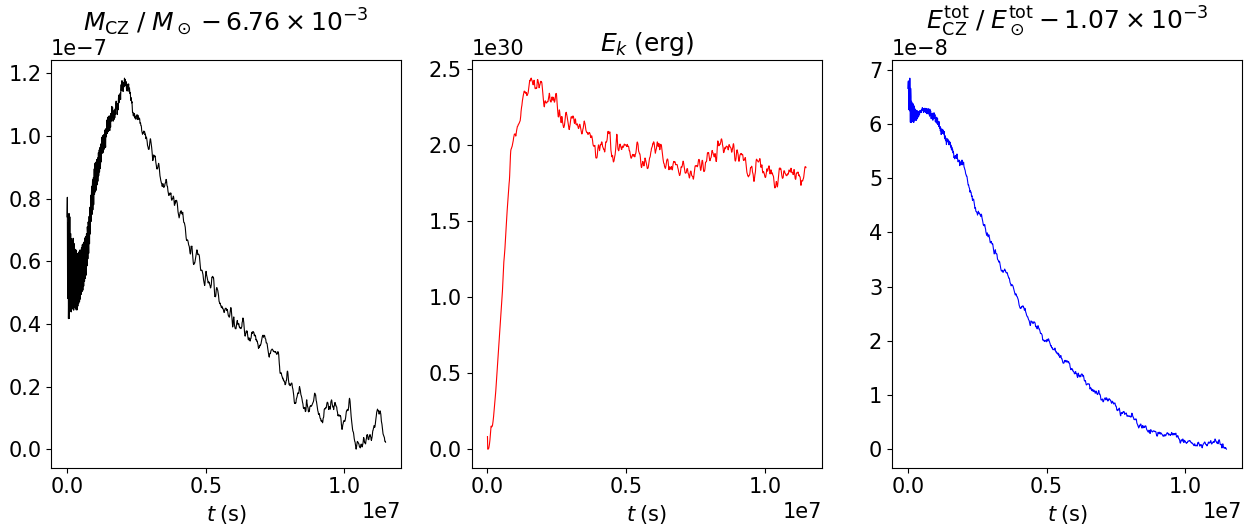}
        \caption{Temporal evolution of integrated quantities in the convection zone (CZ).
        The quantities shown are respectively the total mass in the CZ divided by the total solar mass ($M_\mathrm{CZ} \; / \; M_\odot$), the kinetic energy ($E_k$), total energy in the CZ divided by the total energy in the whole domain ($E_\mathrm{CZ}^\mathrm{tot} \; / \; E_\odot^\mathrm{tot}$). The kinetic energy is saturated, the mass and the total energy are well conserved in the CZ: we observe perturbations of $10^{-7}$ over a mean of $6.76 \times 10^{-3}$ for the mass ratio, and $10^{-8}$ over $1.07 \times 10^{-3}$ for the total energy ratio. }
        \label{fig:conservationtemporel}
    \end{figure}

    The long-term evolution of total mass, kinetic energy, and total energy in the convective zone is tracked to verify the numerical stability of the code for the $Pr=0.1$ model. Despite the non-orthogonal mesh deformations and the large density stratification ($N_\rho \approx 7$), the $\alpha$--$\beta$ well-balanced scheme successfully prevents numerical drift, ensuring that these quantities remain well-conserved over long dynamical timescales (Fig.~\ref{fig:conservationtemporel}). While the kinetic energy reaches full statistical saturation, the total mass and total energy are close to being saturated and exhibit negligible fluctuations.

\section{Validation of the well-balancing implementation}
\begin{figure}
        \centering
        \includegraphics[width=0.9\linewidth]{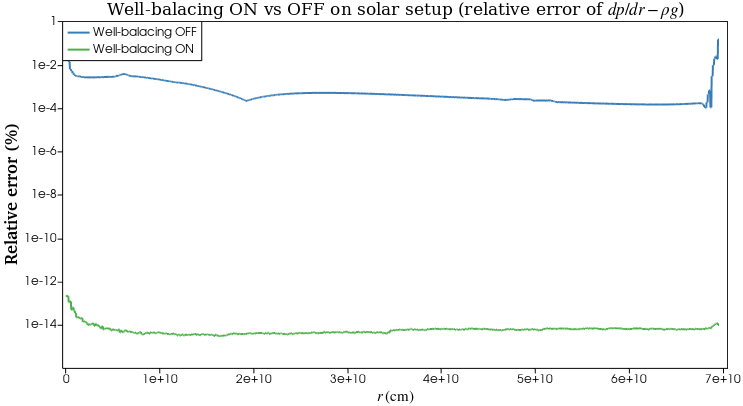}
        \caption{
            Comparison of hydrostatic equilibrium preservation between a standard scheme and the $\alpha-\beta$ well-balancing method after 20 iterations for the global solar setup without any initial perturbation.
            The curves show the relative error of the hydrostatic residue $\dv{p}{r} - \rho g$ averaged over each radius.
            Blue curve: well-balancing disabled. Green curve: well-balancing enabled.
            The $\alpha-\beta$ scheme is 10 orders of magnitude more accurate on the \calhoungrid{}.
        }
        \label{fig:wb_test_hydrostatic}
    \end{figure}

    To verify the accuracy of the well-balancing method (\ref{sec:wellbalacing}), we perform a static numerical test using the fully stratified solar background profile without adding any initial velocity or pressure perturbations. Figure~\ref{fig:wb_test_hydrostatic} shows the radial profile of the relative hydrostatic residue after a few iterations. The $\alpha-\beta$ scheme is 10 orders of magnitude more accurate than the standard scheme on the \calhoungrid{}.

\section{Convergence test on a Sedov blast wave}
\begin{figure}
        \centering
        \includegraphics[width=0.98\linewidth]{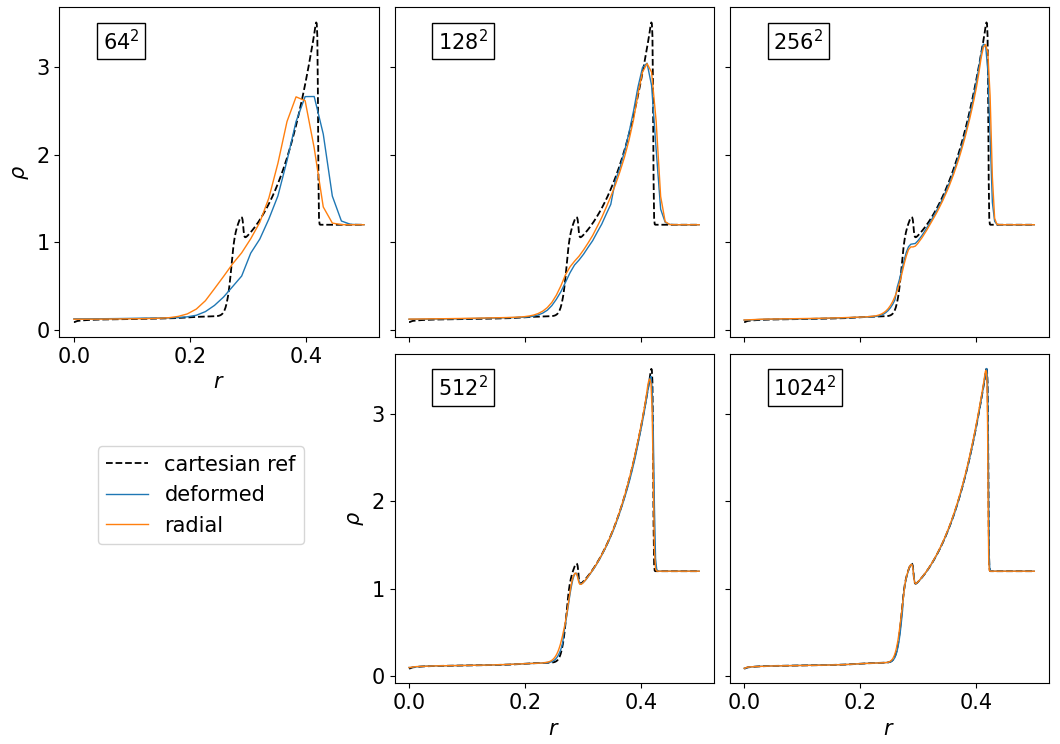}
        
        \caption{
        Convergence test of the Sedov blast wave on radial grid (orange curve) and deformed Cartesian grid (blue curve) with resolutions from $64^2$ to $1024^2$. Test parameters are: adiabatic index $\gamma = 5/3$, initial pressures $p_\mathrm{in} = 10$, $p_\mathrm{out} = 0.1$, initial densities $\rho_\mathrm{in} = 1$, $\rho_\mathrm{out} = 1.2$, and initial blast radius $r_0 = 0.1$. Density profiles are shown at $t=0.2\,\mathrm{s}$, $r=0$ corresponds to the grid center. The reference solution is a $1024^2$ Cartesian grid (black dashed line). The solutions smoothly converge towards the Cartesian reference as the resolution increases, both curves becoming nearly indistinguishable at the highest resolution.
        }
        \label{fig:convergence_blast}   
    \end{figure}

Figure~\ref{fig:convergence_blast} shows a convergence test on the Sedov blast wave problem. We ran this test on the radial and deformed Cartesian grids at different grid resolutions and compared it against a high-resolution reference solution on a $1024^2$ Cartesian grid. As the resolution increases, the solution on both deformed and radial grids converges to the reference Cartesian solution and becomes indistinguishable at sufficiently high resolution. 
\label{appendix:blast}

\section{Solution of the background state Eq.\ref{eq:background_miesch}}
    \label{anx:solve_background}

    In this section we solve the equation for the initial background state starting from the equation \ref{eq:background_miesch}.
    
    By applying the variable substitution $\psi = \rho^{\gamma - 1}$, this equation is transformed into a standard first order linear ODE:
    \begin{equation}
        \dv{\psi}{r} + A(r)\,\psi + B(r) = 0 
    \end{equation}
    where the radial coefficients are defined as $A(r) = \frac{\gamma - 1}{c_p} \dv{S}{r}$ and $B(r) = g(r) \, \frac{\gamma - 1}{\gamma} e^{-\gamma\frac{S(r)}{c_p}}$. 

    This linear equation yields an explicit analytical solution for the radial density profile:
    \begin{equation}
        \rho(r) = 
        e^{-\frac{S(r)}{c_p}} 
        \hspace{-0.2em}\left( \left( 
        \rho_0 e^{\frac{S_0}{c_p}} \right)^{\hspace{-0.3em} \gamma - 1} \hspace{-0.5em} 
        - \frac{\gamma-1}{\gamma}\hspace{-0.2em} 
        \int_r^{R_\odot} g(r') e^{-\frac{S(r')}{c_p}} \dd{r'}
        \right)^{\hspace{-0.5em} \frac{1}{\gamma-1}}
    \end{equation}
    where the integration constants $\rho_0$ and $S_0$ are determined using reference values at the solar surface ($r = R_\odot$). The background pressure $p(r)$ is subsequently recovered by integrating the hydrostatic equilibrium equation, while the entropy profile $S(r)$ is obtained by direct integration of its gradient. The specific heat capacity $c_p$ is adjusted such that the total integrated mass of the generated profile strictly matches the solar mass $M_\odot$ ($c_p \approx 3.25 \times 10^8 \, \mathrm{erg} \cdot \mathrm{g}^{-1} \cdot \mathrm{K}^{-1}$).

    We solve this equation numerically in Python to construct a high-resolution cubic spline representation, which is then exported as a parameter file loadable by \texttt{DWS}. This approach ensures a computationally efficient and highly accurate evaluation of the background states and their spatial derivatives at any arbitrary radial coordinate (e.g., cell centers or face centers), even though the ODE does not have a closed-form solution.

\begin{figure}
        \centering
        \includegraphics[width=0.8\linewidth]{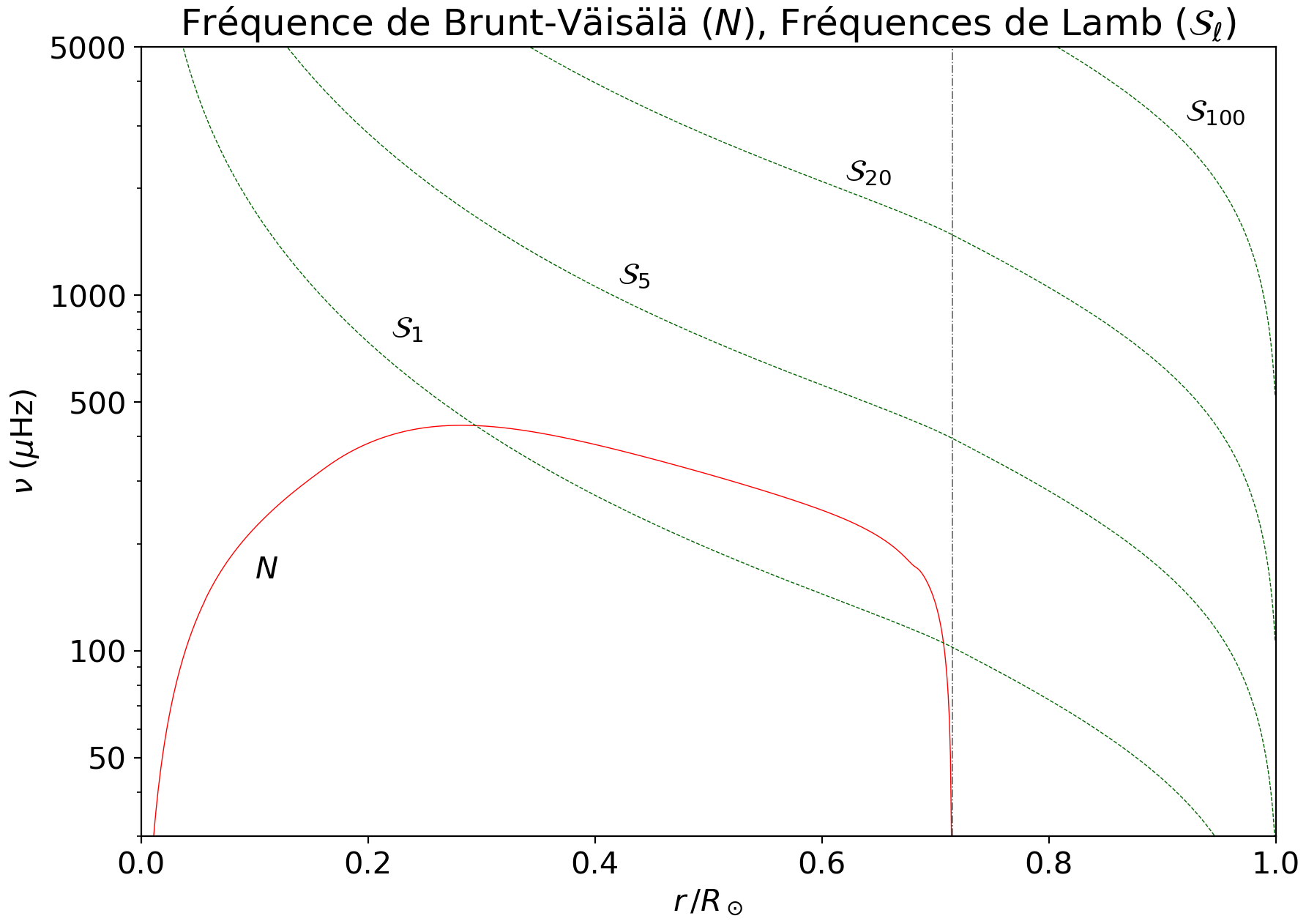}
        \caption{Radial profiles of the Brunt-Väisälä frequency ($N$) and Lamb frequencies ($S_\ell$) for the background state. The former defines the propagation zone for Internal Gravity Waves (IGWs), and the latter for acoustic waves. IGWs propagate only within the stable radiative zone and are evanescent in the convection zone. The black dot--dashed line denote the tachocline. The blue dashed line correspond to a radius of $r=0.32\, R_\odot$.}
        \label{fig:brunt_lamb}
    \end{figure}

    In figure~\ref{fig:brunt_lamb}, we show the profile of Brunt-Väisälä and Lamb frequencies used in the various models that help delimit the resonance cavity for respectively the g and p-modes.

\subsection{Spatial power spectra in the convective zone}
\begin{figure}
        \centering
        \includegraphics[width=0.9\linewidth]{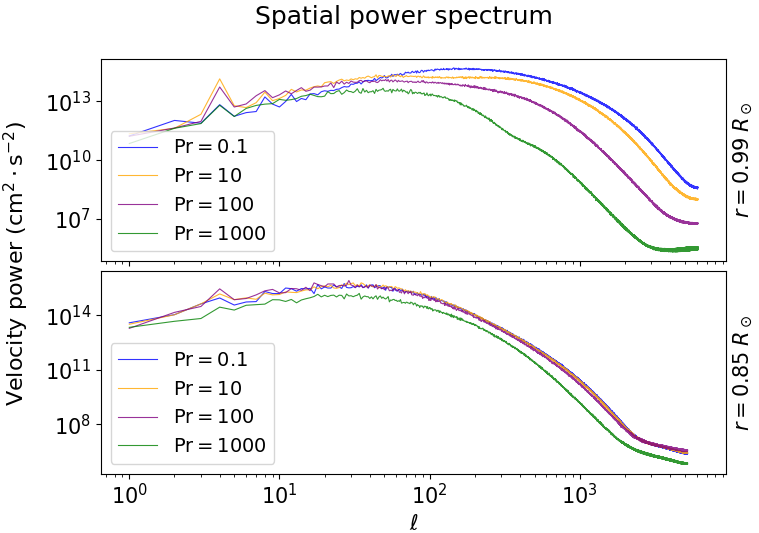}
        \caption{
        Spatial power spectra of radial velocity at $r = 0.99 \, R_\odot$ and $r = 0.85 \, R_\odot$ for different Prandtl numbers. Close to the surface ($r = 0.99 \, R_\odot$), high--Prandtl number runs have less energy at small scales ($\ell \ge 100$) due to strong viscous damping. However, deeper in the CZ ($r = 0.85 \, R_\odot$), the spectra are similar across the different Prandtl numbers because the small-scale downward plumes have already merged into larger ones. 
        The saturation of the spectrum at low Prandtl numbers indicates that the scheme's numerical viscosity dominates below $\mathrm{Pr} \approx 0.1\text{ -- }1$.
        }
        \label{fig:fftspatial}
    \end{figure}

    In complement to the figure~\ref{fig:sun_calhoun_refine_multi_prandtl} showing how the convective pattern varies with the Prandtl number of the simulation, we illustrate in figure \ref{fig:fftspatial} the power spectrum as a function of the spherical harmonics $\ell$. As evident in the physical space the convective pattern near the surface changes drastically with $\mathrm{Pr}$. The case with the lowest Prandtl number corresponding to the most turbulent case is the busiest small scale intense downdrafts. By contrast the case with the largest Prandtl number exhibits broader up and downflows.
    In the spectra, near the surface the difference between the various models is also obvious, we see that the case with the high Prandtl number has the least energy in the small scale (large $\ell$ values). We also note that the peak value changes going toward smaller scale as we decrease $\mathrm{Pr}$. For completeness we also plot the case with the $\mathrm{Pr}=100$ whose spectra is intermediate between $\mathrm{Pr}=10$ and $\mathrm{Pr}=1000$, but we did not illustrate this case in the figure~\ref{fig:sun_calhoun_refine_multi_prandtl}.
    Near the surface, the power spectra saturate at low $\mathrm{Pr}$, showing that numerical viscosity dominates below $\mathrm{Pr} \approx 0.1\text{ -- }1$.
    Deeper in the convection zone, we see that the spectra tends to converge toward similar behavior, in agreement to the convergence of the filling factors reported in the caption of figure~\ref{fig:sun_calhoun_refine_multi_prandtl}.
    
\end{document}